\documentstyle[twocolumn,aps,prb,floats]{revtex}
\input epsf.sty
\newcommand{\be}{\begin{equation}}
\newcommand{\ee}{\end{equation}}
\newcommand{\bea}{\begin{eqnarray}}
\newcommand{\eea}{\end{eqnarray}}
\newcommand{\beas}{\begin{eqnarray*}}
\newcommand{\eeas}{\end{eqnarray*}}
\newcommand{\bfri}{\mbox{${\bf r}_i$}}
\newcommand{\bfrj}{\mbox{${\bf r}_j$}}
\newcommand{\bfr}{ \mbox{${\bf r}$} }
\newcommand{\bfu}{\mbox{${\bf u}$}}
\newcommand{\exc}{ \mbox{$\epsilon_{xc}({\bf r})$} }
\newcommand{\ex}{ \mbox{$\epsilon_{x}({\bf r})$} }
\newcommand{\ec}{ \mbox{$\epsilon_{c}({\bf r})$} }
\newcommand{\nxc}{ \mbox{$n_{xc}$} }
\newcommand{\enxc}{ \mbox{$e_{xc}({\bf r})$} }
\newcommand{\avnxc}{ \mbox{$\langle n_{xc}( {\bf r},u ) \rangle$} }
\newcommand{\avnxcld}{ \mbox{ $\langle n_{xc}^{LDA}( {\bf r},u ) \rangle$ } }
\newcommand{\avnxcwd}{ \mbox{ $\langle n_{xc}^{WDA}( {\bf r},u ) \rangle$ } }
\newcommand{\avnxcad}{ \mbox{ $\langle n_{xc}^{ADA}( {\bf r},u ) \rangle$ } }

\begin{document}
\bibliographystyle{prsty}
\draft
\preprint{\today}
\title{Comparative study of density functional theories of the 
exchange-correlation hole and energy in silicon}
\author{A. C. Cancio and M. Y. Chou}
\address{School of Physics, Georgia Institute of Technology, 
Atlanta GA 30332-0430} 
\author{Randolph Q. Hood}
\address { Lawrence Livermore National
Laboratory, Livermore, CA 94551}
\address{\rm (Submitted to Physical Review B)}
%\date{\today}
%\maketitle
\address{\mbox{ }}
\address{\mbox{ }}
%\begin{abstract}
\address{\parbox{14cm}{\rm \mbox{ }\mbox{ }
We present a detailed study of the exchange-correlation hole and 
exchange-correlation energy per particle in the Si crystal as calculated
by the Variational Monte Carlo method and predicted by various
density functional models.  Nonlocal density averaging methods prove to be 
successful in correcting severe errors in the local
density approximation (LDA) at low densities 
where the density changes dramatically over the correlation length of the
LDA hole, but fail to provide systematic improvements at higher densities where
the effects of density inhomogeneity are more subtle.  
Exchange and correlation considered separately show a sensitivity
to the nonlocal semiconductor crystal environment, particularly
within the Si bond, which is 
not predicted by the nonlocal approaches based on density averaging.
The exchange hole is well described
by a bonding orbital picture, while the correlation hole has a significant
component due to the polarization of the nearby bonds, which partially
screens out the anisotropy in the exchange hole.
%Both are less successfully
%described by the density averaging theories we have considered than the total
%exchange-correlation hole.
}}
\address{\mbox{ }}
\address{\mbox{ }}
%\end{abstract}
\address{\parbox{14cm}{\rm PACS numbers: 
71.15.Mb, 71.45.Gm, 02.70.Ss}}
\maketitle

\makeatletter
\global\@specialpagefalse
\def\@oddhead{REV\TeX{} 3.0\hfill Cancio et al.\ Preprint, 2001}
\let\@evenhead\@oddhead
\makeatother

\narrowtext

\section{Introduction}
%Enables
%comparison with models of $n_{xc}$ used to generate
%nonlocal/semilocal density functional theories (DFT's)
%Identification of problem areas for current models.
%Development of new ideas for nonlocal DFT's.
%``To do surgery, first learn anatomy"
%
%Possibly -- look at the contribution of the exchange-correlation energy
%per particle to the exchange-correlation potential for WDA, ADA and LDA.  
%The WDA and ADA forms are 
%questionable, yet WDA is used to improve the ground-state structural 
%properties of crystals.
%Possibly -- note how the discrepancy of the LDA from VMC data can at least
%partly be understood by a gradient expansion -- if and only if 
%one includes $\nabla^2 n({\bf r})$, unlike standard GGA's.
%
Density functional theory (DFT)
%which models the ground-state
%energy as a functional of the density in the Kohn-Sham formalism, 
is the leading theoretical tool for the study
of the material properties of solids.  
It is based on the characterization of the ground-state energy of the 
inhomogeneous electron gas as a functional of the density,
whose optimization condition can be expressed in terms of the self-consistent
single-particle Kohn-Sham equation.~\cite{KS}
The success of the method lies
in the ease of use and surprising effectiveness of the standard
approximation, the local density approximation (LDA).  This 
models the key component of the functional, the exchange-correlation
energy $E_{xc}[n]$, 
%and its functional derivative with respect to the density, 
and the exchange-correlation potential used in the Kohn-Sham
equation with the assumption that the inhomogeneous electron gas at
any location in space behaves like the homogeneous
electron gas at a density equal to the local density.
As a result, the formidably complex and nonlocal 
relationship between the energy and the density 
due to interparticle correlations in the inhomogeneous environment 
can be converted into a local functional dependent on input 
from the comparatively simpler homogeneous electron gas.  
This approximation has been quite successful at obtaining
useful and suprisingly accurate estimates for bulk ground-state properties
of many solids and the basic ground-state structure of solids, surfaces
and molecules.  

However, there remains a significant need for the development of more
accurate density functionals in several areas.
Applications in molecular and solid-state chemistry require the 
calculation of total energies to at least an order of 
magnitude greater precision than the LDA can provide.  
Excited state properties, in particular the band-gap of semiconductors, 
are not typically obtainable with quantitative accuracy.  
In addition, in materials in
which electron correlations are important, the LDA often fails to predict
qualitative ground-state properties.  
There have been numerous attempts in recent years to develop
exchange-correlation functionals beyond the LDA,
including generalized gradient approximations
(GGA's)~\cite{LM,LYP,BeckeGGA,PW91,PBE}
based on gradient expansions about the LDA,
model nonlocal potentials,~\cite{GJL,AlonsoWDA}
self-interaction corrections,~\cite{PZ} hybrids with 
Hartree-Fock~\cite{Becke} and configuration interaction~\cite{Savin} or
other many-body theories,~\cite{Langreth,KurthP} and orbital-dependent
functionals.~\cite{OEP,KLI,EXXL,EXXB}  
These have led to significant improvements in accuracy
of DFT calculations; but a functional of systematically quantitative
accuracy for many quantum chemistry and solid-state applications has
yet to be achieved.

A key relation that has been particularly
fruitful in the development of density functional theory is the 
adiabatic connection formula~\cite{Exc2nxc} which relates the 
exchange-correlation
energy to a coupling-constant integrated exchange-correlation hole.  
The exchange-correlation hole, $n_{xc}$,
measures the change in density from its mean value at each
point in a system, given the observation of an electron at one particular
point, providing a simple
visualization of the effects of electron correlation in an inhomogeneous
material.
By integrating this quantity over a family of systems characterized
by interaction coupling constant $\lambda$ and kept at fixed density, 
the kinetic energy cost to create the hole at full coupling, $\lambda\! =\! 1$,
is incorporated into it.
As a result, the Coulomb interaction energy of the integrated 
hole reflects the total energy associated with creating the hole.
%both the potential energy gain and kinetic energy cost of creating the hole.
%(For the relation between $E_{xc}$ and the hole to be exact, a 
%potential $V_{\lambda}$ is introduced to keep the ground-state
%density invariant with $\lambda$.) 

The adiabatic connection formula, by making explicit the relation between
the exchange-correlation energy and the exchange-correlation hole has been
the impetus, both explicitly and implicitly, to 
many proposed corrections to the LDA.  
The average density approximation~\cite{GJL} (ADA) and weighted
density approximation~\cite{GJL,AlonsoWDA} (WDA) make explicit use of this
relation to construct nonperturbative nonlocal density functionals.
WDA in particular has attracted attention as a potential tool with
applications for
quantum chemistry,~\cite{WDAChem} metallic surfaces,~\cite{JG} and
bulk solid-state structure.~\cite{Singh,Chuck,Mazin,HybLou}
It is apparently limited by a lack of consistency in its 
behavior~\cite{Singh,Chuck} and typically requires the use of various
context-dependent models or extensions to achieve optimal
results for different applications.~\cite{WDAChem}  
The PW91~\cite{PW91} and PBE~\cite{PBE}
forms of the GGA explicitly build in a number of exact and approximate
properties of the system average of the exchange-correlation hole and 
energy using a gradient expansion of $n_{xc}$~\cite{PBW-GGA} as a framework.  
%However, this
%framework has not been developed as an accurate model
%for a point-by-point analysis of $n_{xc}$ or its associated
%energy density, but rather to reproduce 
%system-averaged quantities such as the total energy.~\cite{PBW-GGA}
%For example, the form for the hole has nonanalytic behavior that is
%ameliorated only in the system average, and 
%and is based on a gradient expansion that eliminates by an integration
%by parts the Laplacian terms that occur in the expansion of $n_{xc}$ 
%-- again correct in the system average.
%Finally, hybrid models combining features of Hartree-Fock or
%configuration interaction theory with density
%functional theory,~\cite{Becke,Savin} which have been particularly 
%successful in achieving quantitative improvements of binding and
%total energies in molecules, can be motivated as an approximation to the
%coupling constant integration of the adiabatic connection formula,
%applying different approximations for different points along the integration
%path.~\cite{Levy}
Hybrid models,~\cite{Becke,Savin} which have been particularly 
successful in achieving quantitative improvements of binding and
total energies in molecules, may also be motivated as approximations to the
coupling constant integration of the adiabatic connection formula.~\cite{Levy}

In developing many of these approaches, many-body calculations of the 
exchange-correlation hole
and related quantities in real systems, particularly
in atoms and simple molecules, have played a useful role,
helping to confirm assumptions of approximate functionals or point out 
specific areas for improvement.~\cite{EPB,BPEontop,Cancio}  
There have been few such studies
for real solids, however.~\cite{Hood1,Hood2}

A second fruitful decomposition of $E_{xc}$ has been to separate it into
exchange and correlation components.  The exchange energy, which contains the
correlation due to Fermi statistics that occurs for the noninteracting
or $\lambda\! =\! 0$ limit, may be explicitly written in terms of the 
occupied single-particle orbitals obtained in this limit, and thus it
and its functional derivative with density can in principle
be obtained exactly and self consistently.  Applications along
this line include the optimized effective potential (OEP) method,~\cite{OEP}
an implementation of the exact exchange potential which
has been limited in practice to free atoms and other
simple geometries, and the
method of Krieger, Li and Iafrate (KLI)~\cite{KLI} which provides an approximate
evaluation of the exchange functional by means of a simplified 
single-particle Green's function.  Recently, an 
exact-exchange formalism applicable to solids has been developed by
St{\"a}dele et al.~\cite{EXXL,EXXB}  These methods are of particular interest 
because they are self-interaction free and are promising candidates for the 
quantitative treatment of excited states.~\cite{EXXB,BylKl}

On the other hand, these advances expose an important problem: many
traditional methods such as LDA rely for their success on the close
relation between exchange and correlation, in which the correlation
hole tends to correct for much of the nonlocal behavior in the exchange 
hole.~\cite{Exc2nxc}
As a result, the error for the sum of exchange and correlation 
energies is typically
significantly smaller than for either separately.~\cite{EBP}  
It is of interest, therefore, to study how the correlation hole alone
behaves in real materials and how current functionals for 
correlation fare with respect to accurate many-body calculations.
Or from the complementary point of view, it becomes necessary to consider
to what extent modeling orbital effects in the correlation hole become 
important once they are included in exchange.~\cite{KurthP}

%INTRODUCTION
%In particular, there has been a growth in the use of explicitly orbital
%dependent density functionals, particularly in the implementation of the
%exchange energy and potential.  
%There is a subsequent need to understand how standard 
%density functional models of correlation hole alone, 
%rather than full exchange-correlation hole hold up against realistic
%numerical data.

%Detailed study of exchange-correlation hole $n_{xc}$ in prototypical 
%semiconductor Si.
%Taking advantage of recent {\it a priori} quantum Monte Carlo (QMC)
%data for $n_{xc}$~\cite{Hood1,Hood2} 

Variational Quantum Monte Carlo (VMC) calculations
of the coupling-constant pair correlation function for the Si crystal have
been reported previously,~\cite{Hood1,Hood2} 
including an analysis in depth of the exchange-correlation energy density 
obtained by the adiabatic connection theory.  
In this work we discuss in more
detail the exchange-correlation hole, its decomposition into
exchange and correlation and the relation between the angle-averaged
hole and the energy density.
We present analysis along two lines: first a detailed 
comparison of the VMC exchange-correlation hole and that of several
DFT models as a function of position in the crystal, in order to search
for systematic trends in how these functionals describe the 
exchange-correlation hole of a prototypical semiconductor material.  
Secondly, we look at
how the exchange and correlation holes separately behave in the
bonding region of Si, and observe how orbital effects in the exchange
generate related effects in the correlation hole.

The paper is organized as follows: Sec.~\ref{theory_nxc} provides 
theoretical background on the exchange-correlation hole, and the various
models used to describe it.  We briefly discuss the method
used to obtain computational results in Sec.~\ref{method}.
Results for spherically averaged holes are presented in 
Sec.~\ref{results_avnxc}, and a detailed analysis of 
the correlation and exchange holes in Sec.~\ref{results_bondcenter}.
In Sec.~\ref{Discussion} we discuss issues raised by our data analysis
concerning the improved treatment of nonlocal density information
in DFT's, as well as orbital effects in the Si correlation and exchange holes
that may be of relevance in improving upon exact exchange methods.
Our conclusions are presented in Sec.~\ref{Conclusion}.

\section{The exchange-correlation hole in density functional theory}
\label{theory_nxc}
\subsection{Basic theory}
We consider a family of systems parameterized by a coupling-constant 
$\lambda$
\begin{equation}
       H(\lambda) = 
         \sum_i \left( -\frac{\nabla_i^2} {2} + V_{ext}({\bf r_i}) 
       + V_{\lambda}({\bf r_i}) \right) \\
       + \sum_{i>j} \frac{\lambda} {|{\bf r_i} - {\bf r_j}|}
       \label{eqHam}
\end{equation}
where $V_{ext}$ is the external potential and $V_{\lambda}$ a potential added
to keep the ground-state density invariant.  The units here and elsewhere
in the paper are in Hartree atomic units unless otherwise indicated.
Two limits of interest include the noninteracting system, $\lambda\! =\! 0$, 
in which $H$ reduces to the standard Kohn-Sham equation, with the
adjustable potential equal to the Kohn-Sham potential. 
%$V_{KS}$.
The second is at $\lambda\! =\! 1$, 
where $V_{\lambda}$ vanishes and one recovers the original
fully interacting many-body system.  The ground-state wave function is
a Slater determinant of single-particle orbitals in the first case and the
true many-body ground state in the second.

The exchange-correlation hole is defined for a given value of $\lambda$ as the 
change in the ground-state expectation of the density at one point in the 
system ${\bf r+u}$ given the observation of an electron at some other point 
${\bf r}$:
\begin{eqnarray}
    \nonumber  n_{xc}& &({\bf r}, {\bf r}+{\bf u};\lambda) = \\
    \nonumber & & \frac{1}{n({\bf r})}\;  
               \langle\, \sum_i \delta({\bf r}-{\bf r}_i) 
                         \sum_{j\neq i} \delta({\bf r}+{\bf u}-{\bf r}_j)\,
	      \rangle_{\lambda} - n({\bf r}+{\bf u}).\\
        & & 
     \label{eqnxc}
\end{eqnarray}
Here $\langle\rangle_{\lambda}$ indicates an average over the ground-state
wave function of $H(\lambda)$ and 
$n({\bf r}) = \langle\, \sum_i \delta({\bf r}-{\bf r}_i) \rangle_{\lambda}$ 
is the $\lambda$-invariant ground-state density.
A related quantity, the pair correlation 
function $g({\bf r},{\bf r}+{\bf u})$, 
is a measure of the pair density relative to that expected for uncorrelated
electrons with the same density distribution.  The
exchange-correlation hole is expressed in terms of $g$ as
\begin{equation}
      \nxc({\bf r},{\bf r}+{\bf u};\lambda) = 
     n({\bf r}+{\bf u})\; [\:g({\bf r},{\bf r}+{\bf u},\lambda) - 1\:].
     \label{eqpcf}
\end{equation}

The importance of the exchange-correlation hole to density functional 
theory lies in its connection~\cite{Exc2nxc,Levy} to the 
exchange-correlation energy $E_{xc}$:  
\begin{equation}
   E_{xc}[n] = \frac{1}{2}\int\!d^3r\: n({\bf r}) \int\!d^3r^{\prime} 
   \int_0^1 \!d\lambda\: 
     \frac{\displaystyle n_{xc}({\bf r},{\bf r}^{\prime},\lambda)}
      {\displaystyle  |{\bf r} - {\bf r}^{\prime}|}.
   \label{eqexctot}
\end{equation}
$E_{xc}$ includes all the contributions to the energy due to correlations,
that is, beyond the noninteracting and Hartree energy.
The exchange-correlation energy $E_{xc}[n]$ is 
analyzed in density functional
theory in terms of the exchange-correlation energy density $e_{xc}({\bf r})$
and the exchange-correlation energy per particle $\epsilon_{xc}({\bf r})$
\begin{equation}
  E_{xc} = \int\! d^3r\; n(r) \epsilon_{xc}({\bf r}) 
	       = \int \!d^3r\; e_{xc}({\bf r}).
   \label{eqexc}
\end{equation}
Relating this expression to that for $E_{xc}$ in terms of the 
exchange-correlation hole $n_{xc}$ one has
\begin{equation}
 \epsilon_{xc}({\bf r}) = \frac{1}{2} \int\!d^3u
     \frac{\displaystyle n_{xc}( {\bf r}, {\bf r}+{\bf u} ) }
      {\displaystyle u},
   \label{eqexc2}
\end{equation}
where $n_{xc}( {\bf r}, {\bf r}+{\bf u} )$ is the coupling-constant 
integrated hole,
$\int_0^1\! d\lambda\: n_{xc}({\bf r},{\bf r}+{\bf u},\lambda)$.
%\begin{equation}
%     n_{xc}( {\bf r}, {\bf r}+{\bf u} ) = \int_0^1 \!d\lambda\: 
%                n_{xc}({\bf r},{\bf r}+{\bf u},\lambda).
%\end{equation}
The exchange-correlation energy per particle thus has the natural 
interpretation as the interaction energy of the particle with its 
exchange-correlation hole.  

The exchange-correlation hole and energy are frequently decomposed according
to separate exchange and correlation components.  The exchange hole $n_x$
is that of the noninteracting or $\lambda\! =\! 0$
ground state [Eq.~(\ref{eqnxc})] and contains correlations
due solely to Fermi statistics.  The remainder, $n_c$,
of the contributions to the total $n_{xc}$ are those induced by 
introducing pair correlations into the ground-state wave function,
representing density fluctuations caused by the Coulomb interaction.
The two components satisfy important sum rules which reflect the conservation
of particle number:
\begin{eqnarray}
    \displaystyle\int \!d^3u\; n_x({\bf r},{\bf r}+{\bf u}) &=& -1, \\
    \displaystyle\int \!d^3u\; n_c({\bf r},{\bf r}+{\bf u}) &=& 0. 
\end{eqnarray}
%Essentially, the exchange hole sum rule indicates the removal
%of the electron observed at ${\bf r}$ from the density observed elsewhere,
%while the correlation hole measures fluctuations caused by the Coulomb 
%interaction in the density of the remaining electrons and produces
%no net change in their number.  The sum rule
%of $\nxc$ is simply derived from the sum of its two components.

%The exchange-correlation hole enters into DFT in another important way
%by determining the exchange-correlation potential $v_{xc}({\bf r})$ used
%in the Kohn-Sham equation for the density.  This equation is essentially
%the requirement that the functional derivative of the total energy should be 
%zero for the ground state density.  The exchange-correlation potential is
%then the contribution of $E_{xc}$ to this condition:
%\begin{equation}
%    v_{xc}({\bf r}) = \frac{\delta E_{xc} }{ \delta n({\bf r}) }.
%\end{equation}

A key to the practical exploitation of the relation between $n_{xc}$ 
and $E_{xc}$ is that, since the Coulomb interaction depends only on
interparticle distance, much of the information contained in $n_{xc}$ is
not needed and may be ignored in the development of models.
In particular, to obtain the energy per particle $\exc$ at a given 
point ${\bf r}$ the angular variation in $n_{xc}$ may be integrated out 
to leave a function depending only on the distance $u$ from ${\bf r}$,
\begin{equation}
    \avnxc = \displaystyle\int \frac{d\Omega_{u}}{4\pi} 
     n_{xc}( {\bf r},{\bf r}+{\bf u} ),
    \label{eqavnxc}
\end{equation}
with \exc in Eq.~\ref{eqexc2} given in terms of the integral over $u$ of the
weighted angle-averaged hole, $2\pi u \avnxc$.
%\begin{equation}
%    \exc = 2\pi \displaystyle\int du \;u \avnxc.
%    \label{eqavexc}
%\end{equation}
%The weighted, angle-averaged hole, $2\pi u \avnxc$ then gives a picture
%of the hole about a given position which emphasizes the region which 
%contributes most to the exchange-correlation energy
%at that point.  
The DFT models we study in this paper can be viewed
as models of this angle-averaged hole.

\subsection{Models for the exchange-correlation hole} 

%The local density approximation (LDA) may be obtained by  
The LDA model for $n_{xc}$ may be obtained by  
replacing the true hole about ${\bf r}$ with that of the
homogeneous electron gas at the local density, $n({\bf r})$:
        \begin{equation}
           \avnxcld = n({\bf r}) \{ g^{heg}[u,r_s({\bf r})] - 1 \}.
	\end{equation}
The factor $r_s = (3/4\pi n)^{1/3}$ is the average distance between 
electrons in the homogeneous gas of density $n$.  
It provides a useful measure of the correlation
length for the exchange-correlation hole, as the width of the
exchange hole scales linearly with $r_s$ and that of the correlation
hole shows similar behavior for the range of densities $1.4 < r_s < 4.4$
valid for the Si crystal valence.~\cite{OrtizB}

%The weighted density approximation~\cite{JG} attempts to account for the
The WDA~\cite{JG} attempts to account for the
inhomogeneity of the density in realistic systems by incorporating it 
in the density prefactor of $n_{xc}$ 
%(compare Eq.~\ref{eqpcf}):
	\begin{equation}
           \avnxcwd = \displaystyle\int \frac{d\Omega_{u}}{4\pi} 
		n({\bf r}+{\bf u})
		  \{ g^{model}[ u,\bar{r}_s({\bf r}) ] - 1 \}.
	   \label{eqwda}
	\end{equation}
This form of the exchange-correlation hole treats the
hole at ${\bf r} + {\bf u}$ as the modification of the density at that 
point rather than that at the center of the hole. 
This is a property of the exact hole
that leads to a significant departure from the homogeneous 
case when the density varies considerably within the effective range 
of the hole.
%based on the reasonable assumption 
%that the effects of inhomogeneity on $n_{xc}$ are primarily
%to be found in the density prefactor of Eq.~(\ref{eqpcf}) rather 
%than in the pair correlation function.  
%Otherwise $n_{xc}$ can take on unphysical values when
%the density changes dramatically over the length scale of the hole.
%$g$ is a model function depending only on interparticle distance and
%an average length scale $\bar{r}_s$.
%A model for a spherically-averaged pair correlation function,
%$\langle g(u,{\bf r}) \rangle$, such as is used above [Eq.~(\ref{eqwda})],
%is sufficient in principle to obtain the exact 
%angle-averaged exchange-correlation hole.  
The major approximation made is
the use of a scaling form for the pair correlation function
where for each $\bfr$ an averaged length scale $r_s({\bf r})$ is determined 
to select $g$ from a family of suitably chosen pair correlation 
functions $g(u,r_s)$.
This length scale is fixed at each point fixed by the particle sum rule
	     \begin{equation}
%                \displaystyle\int 4\pi u^2du \avnxcwd = 
                \displaystyle\int d^3u \: n({\bf r}+{\bf u})
		  \{ g^{model}[ u,\bar{r}_s({\bf r}) ] - 1 \} = -1.
		\label{eqwdasum}
	     \end{equation}
A natural choice for $g(u,r_s)$ in an extended system 
is that of the homogeneous electron gas; this form will be 
considered in this paper.  
%(Another frequently used form is that of
%Gunnarsson et al.~\cite{GJL} which is tailored to reproduce the 
%asymptotic image charge for a metal surface 
%but sacrifices the correct form for $g$ in the limit of slowly varying
%density which is retained by using the homogeneous electron gas form.)
The WDA scaling approximation has the ability to fit certain
properties of the true $n_{xc}$ unattainable by the LDA; in particular
%it reproduces the exact $n_{xc}$ for a single particle system such 
%as the Hydrogen atom.  Perhaps of more relevance in the current context is that
it significantly improves on the description of the exchange-correlation
energy per particle of atoms when a single
electron is removed, obtaining the correct limiting value of 
$1/2R$ versus distance from the atom $R$.~\cite{GJL}
%The WDA in this scaling approximation has the flexibility
%to reproduce the exact 
%exchange-correlation hole in the single electron problem, e.g.\ the H atom,
%with the choice of ${\bar r_s}=\infty$, essentially equating one electron
%with infinite interelectron separation.  A similar situation applies
%in many-body systems when an electron is moved far from the other 
%electrons, as in the asymptotic removal of an electron from an atom.
%It can be shown that the exchange-correlation energy per particle 
%in this case has the correct limiting value 
%the atom $R$.~\cite{}  

%    (These two features have the benefit of eliminating the correlation
%    self-interaction in the limit of a one electron system, providing 
%    an exact density functional theory in this limit.  
%    A motivation for this situation originally was the better treatment of
%    metal surfaces in which the model $g$ could be fixed to give an 
%    accurate description of the image charge.
%    However, the choice of the homogeneous electron gas $g$ has the 
%    potential of describing accurately the situation of a slowly varying
%    density.  This may be a more important limit for realistic calculations.)

The ADA~\cite{GJL} is another method to incorporate into
a model of the exchange-correlation hole information about the nonlocal 
density variation in the vicinity of the hole.  It preserves the 
homogeneous gas model for $n_{xc}$ but determines the $r_s$ factor at
which it is evaluated from an average of the density 
in the neighborhood of the reference point:
	\begin{equation}
           \avnxcad = \bar{n}({\bf r}) \{ g^{heg}[u,\bar{r}_s({\bf r})] - 1 \}
	\label{eqada}
	\end{equation}
with
	\begin{equation}
            \bar{n}({\bf r})  = 
	    \displaystyle\int d^3r' n( {\bf r}^{\prime} ) \,
		w( |{\bf r}\!-\!{\bf r}^{\prime}| , {\bar r_s} ).
	\end{equation}
  Here $\bar{r}_s = ( 3/4\pi \bar{n} )^{1/3}$ and 
  $w$ is a normalized weighting factor.
  The weight $w$ is chosen to obtain the correct $E_{xc}$ in the linear
  response limit
  %, ie.\ for small amplitude density variations 
  % about a homogeneous density,
  and has an effective range set by ${\bar r_s}$.
%The resulting weight has the intuitively sensible property of cutting off 
%the averaging of the density self consistently with the length scale of the
%hole itself.
 The averaging procedure can in this sense be understood as 
  the selecting of a physically reasonable correlation length for
  the hole from an average of the density over the physical range of the
  hole, particularly of value in cases where the density varies rapidly 
  within this range.
%  This should play a role most obviously where the
%  local density provides an poor estimate of the actual correlation length,
%  such as at extrema in the density.
  Unlike the WDA, the ADA does not try to provide a correction to the shape of
  the $n_{xc}$.

    Note that with the use of a radially isotropic
    $g$, only a spherically averaged density enters into the definition of,
    and sum rule for, the WDA exchange-correlation hole:
     \begin{equation}
                \avnxcwd = \langle n({\bf r},u)\rangle 
		     \{ g^{heg}[ u,\bar{r}_s({\bf r}) ] - 1 \},
	 \label{eqavwda}
     \end{equation}
    and
     \begin{equation}
           \langle n({\bf r},u)\rangle = 
		\displaystyle\int \frac{d\Omega_{u}}{4\pi} 
	           n({\bf r}+{\bf u}).
     \label{eqnav}
     \end{equation}
A similar result holds for the ADA, as the density averaging function $w$
is a function only of interparticle distance.
This averaging reduces the information about the nonlocal density 
actually used by these methods from 
the full complexity of the density variation about 
each reference point to a simpler and often slowly 
varying radial function $\langle n({\bf r},u)\rangle$.  
Knowledge of this function
at each point in space, along with the model used for $g$ in the
WDA and $w$ in the ADA, specifies $E_{xc}$ within each approximation.  
    
  Another line of attack in improving the exchange-correlation hole and
  energy, the GGA, has been to design 
  corrections to the LDA hole using information
  about system inhomogeneity contained in the local density gradient.
  One implementation of this approach~\cite{PBE} has led to the 
  construction of a GGA model for $n_{xc}$.~\cite{PBW-GGA}  
  In this and other GGA models the measure of inhomogeneity is the absolute 
  value of the gradient, $|\nabla n({\bf r})| / n({\bf r})$.
  Second-order effects proportional to the Laplacian of the density,
  $\nabla^2 n({\bf r}) / n({\bf r})$, are mapped to a gradient form by an 
  integration by parts.
  This transformation has no effect on system-averaged quantities,
  namely $E_{xc}$ and the average over $\bfr$ of $n_{xc}$.
  However, the local quantities $\nxc(\bfr,\bfr+\bfu)$ and 
  \exc in the GGA are as a result no longer defined in terms of the 
  adiabatic formulae of Eqs.~[\ref{eqnxc}] and~[\ref{eqexc2}] 
  and not directly comparable with our VMC data.~\cite{BCL}

\section{System and computational details}
\label{method}
A description of the computational method used in obtaining numerical many-body
expectations of the pair correlation function and related quantities 
in Si has been presented in a previous paper.~\cite{Hood2}
We present here a short summary of the details relevant to our discussion.

For $V_{ext}$ in Eq.~(\ref{eqHam}) we use 
a norm conserving nonlocal LDA pseudopotential which replaces
the electrons in the atom core.  The system and expectations involve valence 
electrons and their correlations.  
A simulation cell consisting of a $3 \times 3 \times 3$ lattice of
fcc primitive cells of the diamond lattice containing 216 valence electrons
is used in obtaining expectations.  
%(Note that the supercell is needed because exchange-correlation hole as a 
%function of one electron coordinate only lacks translational symmetry 
%of the crystal.)
Finite size effects in the electron-electron interaction were taken into
account by replacing the Coulomb interaction in Eq.~(\ref{eqHam}) with a
truncated potential into which the long-range
Hartree interaction is incorporated.  This form has proved to be successful 
in reducing finite size errors in the Coulomb
energy as compared to the conventional Ewald interaction.~\cite{Williamson}

To describe the ground state for a given
value of coupling constant, a Slater-Jastrow type
wave function was used, consisting of a product of Slater determinants
multiplied by a trial many-body correlation factor: 
    \be
       \psi^{\lambda} = \exp \left( 
	    - \sum_{i\langle j} u^{\lambda}(|\bfri - \bfrj|) + 
	      \sum_i \chi^{\lambda}(\bfri) 
			  \right) D\uparrow D\downarrow.
    \ee
The orbitals used in the Slater determinants for up and down spins
$D_{\sigma}$ are determined from an LDA
calculation and include all the valence-band Bloch orbitals of the
primitive cell periodic on the supercell.  The Jastrow factor introduces
variationally explicit electron-electron correlations with one-body 
corrections to optimize the density.  It is described in detail in the 
paper by Hood et al.~\cite{Hood2}
A variational ground-state wave function optimized over
22 variational parameters was obtained for five 
different values of the coupling-constant, and
recovered 88$\%$ of the correlation energy as compared to nearly exact 
diffusion Monte Carlo calculations for the same system.  
%(Note -- Randy's published
%estimate of 85$\%$ does not agree with his published table values!)

The pair correlation function 
$g( {\bf r}, {\bf r}^{\prime})$ [Eq.(\ref{eqpcf})] was
obtained from the ground-state wave function for each $\lambda$
in order to obtain coupling-constant integrated exchange-correlation holes 
and energies.  
The pair correlation function was expanded in 
symmetrized plane-waves, taking full advantage of the symmetries of the
pair correlation function in the crystal to reduce the basis set to a
reasonable number.
Expectations for the coefficients for symmetrized plane-waves were evaluated
statistically using the Monte Carlo method.  
The resulting data are limited by plane-wave cutoff and statistical errors
which can be checked by comparing the numerically calculated and exact
values of the exchange hole.  The resulting variation in $g_x$ was between
1\% and 6\%. 
The error due to basis set cutoff is most noticeable near the atom core where,
despite the pseudopotential treatment, the rapid variation of the density 
and the valence orbitals leads to large wave-numbers in the exchange hole.

\section{Angle-averaged exchange-correlation hole}
\label{results_avnxc}
\begin{figure*}[p]
%Fig. 1
%\epsfysize=5.8cm
\hskip 2.0cm
\epsfbox{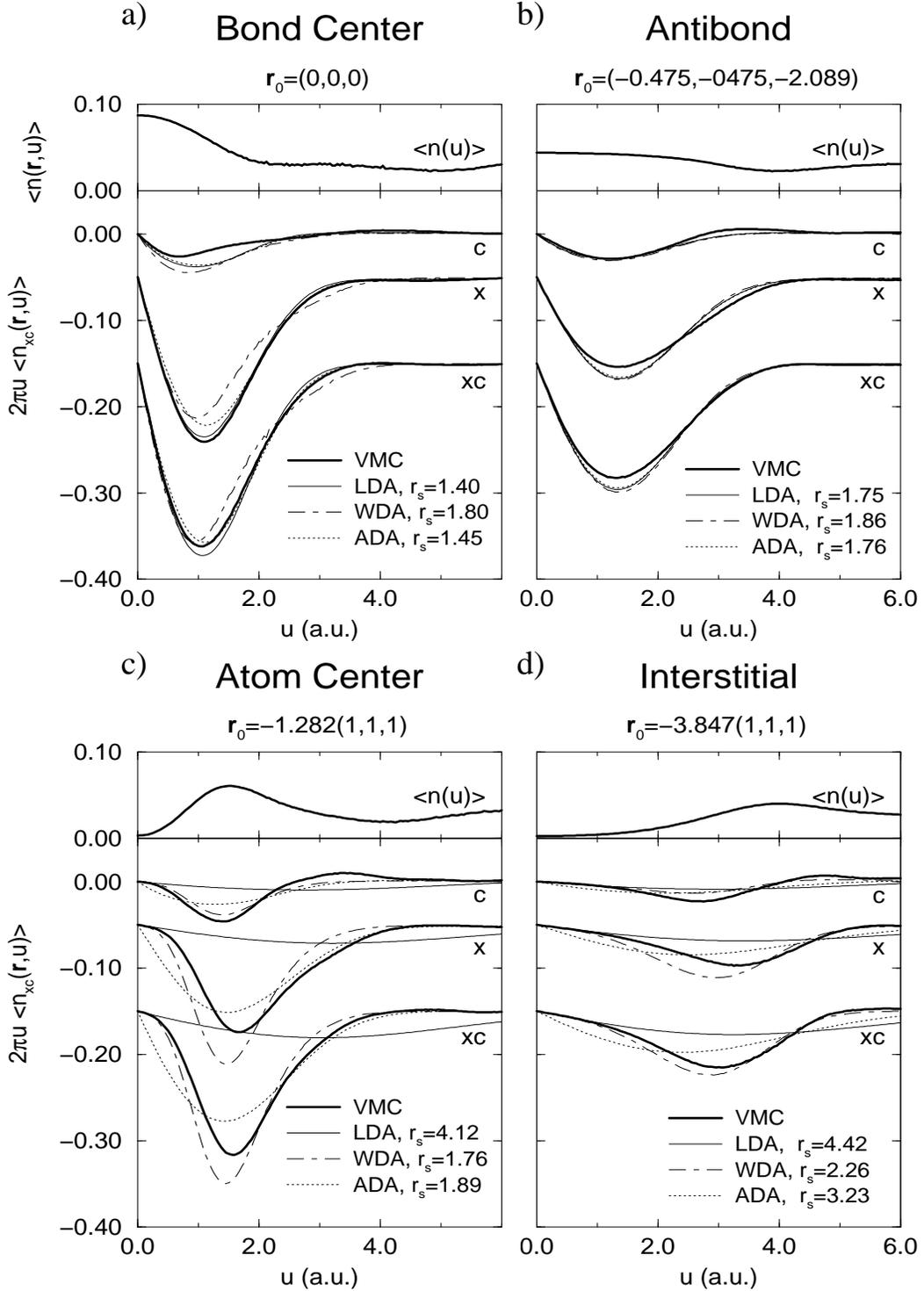}
\vskip 0.3cm
\widetext
\caption
{Distance-weighted angle-averaged exchange-correlation holes at various 
points in the
Si crystal: (a) bond center, (b) antibond position, (c) atom center and (d)
tetrahedral interstitial site in the (110) plane.  Correlation, exchange
and exchange-correlation components for VMC data (thick solid lines) and
several models are shown, with the exchange set off 
by -0.05~a.u.\ and exchange-correlation by -0.15~a.u.\ on the vertical
axis for clarity.
Solid line on the top half of each subplot shows the angle-averaged density
$langle n({\bf r},u)\rangle$ as function of distance from reference electron.
All quantities plotted are in atomic units.
}
\narrowtext
\label{nxcfig1}
\end{figure*}

Figure~\ref{nxcfig1} shows as a function of interelectron distance $u$ the 
angle-averaged correlation,
exchange and exchange-correlation holes, weighted by \mbox{$2\pi u$},
for the VMC calculation of Hood et al.~\cite{Hood1,Hood2} and 
several DFT models at several representative points
in the Si crystal.  
The pair correlation function of the homogeneous electron gas used 
to generate the DFT holes is obtained from an accurate model
form.~\cite{PW}
%The quantity shown is given
%by $2\pi u \langle n_{xc}({\bf r},u)\rangle$, with equivalent expressions for 
%exchange and correlation, where ${\bf r}$
%is the location of a reference electron and $u$ is the distance from the
%reference point.  
%The exchange and exchange-correlation holes have been offset 
%along the vertical axis for clarity; note that the weighted hole in 
%either case should
%exactly equal zero at zero interparticle distance ($u=0$).
The distance weighting factor $2\pi u$ is chosen so that the
integral of the plotted function yields the exchange-correlation energy
per particle at the reference point. 
%as per Eq.~(\ref{eqavexc}).  
These
values are listed separately in Table~\ref{table1}.
%The reference points chosen are at the bond center~(a), 
%near the antibond point on the (111) axis~(b), in the center of the atom 
%core~(c) and at the tetrahedral interstitial point~(d).   

%%%%%%%%%%%%%%%%%% EPS INCLUDE FIG 1, TABLE 1 %%%%%%%%%%%%%%%%%%%%%%%

\begin{table*}[htb]
\widetext
\caption{Exchange-correlation energy per particle $\exc$ in hartees at various
points in the Si crystal discussed in the text.  Label VMC refers to raw
data, VMC-I includes a correction to obtain the correct particle 
sum rule, VMC-II includes a correction for the plane-wave cutoff (see text)
%as estimated 
%by the difference between the exact and plane-wave expanded exchange hole
and corresponds to the integral over the angle-averaged holes shown
in Fig.~\protect\ref{nxcfig1}.
}\label{table1}
\begin{tabular}{l|rrrrrr}
\mbox{$\epsilon_{xc}$} & & & & & & \\
  & VMC & VMC-I & VMC-II & LDA & WDA & ADA \\
\tableline
Bond Center & -0.3743 & -0.3706 & -0.3724 & -0.3804 & -0.3657 & -0.3671 \\    
Bond Axis Right & -0.3427 & -0.3425 & -0.3357 & -0.3803 & -0.3657 & -0.3671 \\
Antibond Point & -0.2947 & -0.2954 & -0.2957 & -0.3090 & -0.3117 & -0.3068 \\ 
Atom Center & -0.2574 & -0.2557 & -0.2783 & -0.1432 & -0.3011 & -0.2885 \\
Interstitial & -0.1654 & -0.1655 & -0.1650 & -0.1348 & -0.1801 & -0.1772 \\
\tableline
\mbox{$\epsilon_{c}$} & & & & & & \\
  & & & VMC-II & LDA & WDA & ADA \\
\tableline
Bond Center &  &  & -0.0317 & -0.0525 & -0.0679 & -0.0513 \\    
Bond Axis Right & & & -0.0374 & -0.0504 & -0.0610 & -0.0500 \\    
Off Bond Axis  & & & -0.0317 & -0.0489 & -0.0585 & -0.0494 \\    
Antibond Point & & & -0.0389 & -0.0478 & -0.0551 & -0.0474     
\end{tabular}
\narrowtext
\end{table*}
%%%%%%%%%%%%%%%%%% EPS INCLUDE FIG 1, TABLE 1 %%%%%%%%%%%%%%%%%%%%%%%

%(Since the ADA weighting function $w$ and the pair 
%correlation function $g$ used in WDA are both functions of distance only,
%they are sensitive only to the average of the density as a function of 
%distance from the reference point about which the hole is constructed.  
%Thus, in both ADA and WDA, all the information about the nonlocal 
%environment that is input into the construction of the exchange-correlation
%hole is contained in this angle-averaged density 
%$\langle n({\bf r}, u) \rangle$.)

In order to give an idea of the
environment surrounding each reference point and to compare with the 
WDA and ADA forms for $n_{xc}$, the angle-averaged density
\mbox{$\langle n({\bf r},u) \rangle$},
the average density at a distance $u$ from the reference point ${\bf r}$,
is shown at the top of each plot.  
%Note that 
%the density is not weighted by distance so that the units of this curve
%are $a_B^{-3}$, versus $a_B^{-2}$ for the other curves.  
The density
is to be compared to the average over the unit cell of 0.0296~a.u.
The angle-averaged density for each of the four reference points shown
indicate four different environments typical of the pseudopotential
model Si crystal.  In the first, (a), the reference electron
is at a density considerably higher than the unit-cell average, and 
the density decreases significantly over the length-scale of the VMC
exchange-correlation hole.  The antibond position (b) is typical of
intermediate density points where the local density is close to the unit-cell
average and the angle-averaged density in the vicinity of the reference point
shows only modest variations over long distances.  The atom center (c)
is typical of the ``pathological" situation inside the pseudopotential core.
Here the density increases by an order of magnitude, from roughly 10$\%$ 
to 200$\%$ of the unit-cell average, over roughly one-third
of the Si bond length of 4.44~a.u., before slowly decaying back to the
average.  Such an environment of dramatic density change over the length
scale of the hole indicates the likely failure of the LDA in this
region, which has in fact been observed in the comparison of LDA and
VMC energy densities.~\cite{Hood1}  Finally, the interstitial point (d) shows
the generic features at low density outside both valence and core.  Here the
angle-averaged density also changes considerably from the local density but 
at a much more gradual rate.  

In the high-density case of the bond center, Fig.~\ref{nxcfig1}(a),
the exchange-correlation hole closely follows that determined by the local
density, with $r_s\!=\!1.397$~a.u.  
%The minimum of the distance-weighted 
%hole is moved slightly towards the reference electron, and at long range
%decays to an energetically insignificant value
%more slowly than the LDA.
%These changes are also present in the nonlocal theories.
The average density within the effective range of the hole
is lower than the local density, leading to a slightly larger ${\bar r_s}$ 
for the ADA and an overall better fit of the hole.  
%The net effect of the lower density outside the immediate vicinity of the 
%fixed electron is somewhat muted by the property of the ADA averaging
%function $w$ which has a selfconsistent cutoff proportional to ${\bar r_s}$.  
The WDA hole [Eq.~(\ref{eqavwda})] shows the effect of two competing factors.
The value of ${\bar r_s}$ used in the scaled pair correlation 
function $g^{heg}$ 
is significantly larger than the local $r_s$. 
%(Note that the sum rule used to obtain
%${\bar r_s}$ weights the exchange-correlation hole by $u^2$ and thus is
%particularly sensitive to the lower average density far from the hole origin.)
The resulting pair correlation function is wider and deeper than that 
of the LDA.  
However, the density prefactor $\langle n(u) \rangle$ in Eq.~(\ref{eqavwda})
drops rather quickly and 
``quenches" the exchange-correlation hole at larger $u$.  Eventually,
as the density settles to its asymptotic value
of the unit cell average, the hole has a long-range tail
again dominated by the shape of the pair correlation function.  
The net result is a considerable overcorrection of the LDA model
with respect to the VMC data.
%The result 
%of these several features is a hole with the peak contribution to 
%the energy shifted towards 
%the core and a total range larger than in the LDA.  On both 
%counts it considerably overcorrects the LDA model with 
%respect to the VMC data.

%(Note that differences between DFT models and VMC are correctly considered
%as deviations rather than errors because the VMC data are neither
%numerically nor variationally exact.  The deviations are significant in that
%the DFT models are heuristic universal functionals of the 
%density while the VMC is a highly accurate many-body calculation of the 
%expectations specifically of the Si crystal system.  The latter therefore 
%can be expected to be a reasonably reliable guide to the exact 
%ground-state expectations of the Si 
%crystal and an indicator of the quality of the universal functionals
%for this specific case.)

At the antibond point, the hole is wider and less deep than at the
bond center, consistent with the variation of the homogeneous
electron gas hole with respect to density.  
Neither form of nonlocal averaging results
in a significant change from the LDA form of $\nxc$ since the angle-averaged
density is almost constant in the vicinity of the reference electron.
The exchange-correlation energy per particle, 
determined from the integral of the curves in Fig.~\ref{nxcfig1}, 
has roughly the same modest deviation 
from the VMC data in each case, as shown in Table \ref{table1}.
This leads to a significant difference in the energy density 
because the density is still large at this point.

The holes at the bond center and antibond point are fairly representative
of those at other points of intermediate-to-high density in the Si crystal.
The corrections introduced by the nonlocal DFT's are quite small because
of the relatively small variation in the angle-averaged density
within the effective range of the hole.  
These high density corrections to $n_{xc}$ and $e_{xc}$ 
have a sensitive dependence on position in 
the crystal, which appears to lack an obvious relation to the 
positional trends in the VMC data.
%These deviations are not smooth functions of the density
%or, for example, density gradient
%but vary significantly from point to point in the crystal.  
As a result, plots of the deviation of the ADA and WDA energy density from 
the VMC data feature quite complex ``dapple" patterns that remain after the 
worst errors of the LDA model are smoothed out.~\cite{Hood1,Hood2}  
%In Sec.~\ref{results_bondcenter}
%we will study more closely the variation of $\enxc$ and $\nxc$ with position
%and how a nonlocal dependence on detailed orbital information beyond the 
%level of the WDA and ADA may be used to explain the observed features.

For the two low-density cases, however, the nonlocal density averaging
techniques have a considerable impact on the shape and quality 
of the exchange-correlation hole.
In these cases the shape of hole is dominated by the 
rapid increase of the density with $u$.  
%In contrast to the situation in the homogeneous electron gas,
%the unweighted hole has a local maximum near $u\!=\!0$, becoming deeper 
%for increasing $u$ because the average density increases 
%and consequently the total density displacement by exchange and correlation 
%increases as well.  In the weighted 
%holes plotted here, this results in a negative second derivative at small
%distances, versus a positive one at higher densities and in the
%homogeneous electron gas.  

At the atom center, Fig~\ref{nxcfig1}(c), the rapid change in density
leads to a far more compact exchange-correlation hole than that
predicted by the LDA at the low local density ($r_s$ = 4.12~a.u.).  
The minimum in the weighted hole, where it contributes the most to the
exchange-correlation energy, is 1.6~a.u.\ from the reference point, less
than half that of the LDA hole.  Roughly $90\%$ of the total contribution
to $\exc$ comes from within a radius of 3.0~a.u., significantly less
than a bond length (4.44~a.u.).  In this case, both ADA and WDA do 
significantly better at determining the overall depth and length scale
of the hole over the energetically important region of 0.5 to 3.5~a.u.
In addition, WDA clearly matches the shape of the VMC hole quite closely.
As a result, the ADA and WDA obtain fairly accurate estimates of the
energy of the exchange-correlation hole as shown in Table~\ref{table1}, 
while LDA obtains only half the VMC value.  

The interstitial point, Fig.~\ref{nxcfig1}(d), has low-density features 
qualitatively similar to the atom center but, consistent with a more
gradual change in density, is considerably shallower and more spread out.
The WDA
has a particularly good overall agreement with the VMC hole on a 
point-by-point basis (although the ADA has a total energy slightly closer to
the VMC value).  However, possibly due to the more gradual change
in density within the vicinity of the hole, the LDA does a much better
job of matching the energy of the hole than near the atom center.  
%(Note, for example, that the minimum of the hole is at 3.17~a.u., as 
%compared to 3.42~a.u.\ in the LDA.)  
The large differences in the shape of the hole largely
cancel out in the integral so that the energy of the hole is only 20$\%$
smaller than the VMC value.  The error in the LDA energy density,
which is obtained by weighting the energy of the LDA exchange-correlation hole 
by the density, is thus fairly insignificant at this point;~\cite{Hood1} 
in contrast the LDA energy density in the atom core 
suffers from considerably larger deviations with the VMC.

Figure~\ref{nxcfig1} also shows results for angle-averaged correlation
and exchange holes at each point.  In general, the exchange hole
is the dominant contribution to the total hole.  The correlation hole
provides a correction to the exchange
hole, reducing electron density near the reference electron and
enhancing it towards the outside of the exchange hole, thereby 
making the total hole slightly deeper and more compact.  At high densities
this correlation hole is weaker and shorter-ranged than the LDA hole,
a result in agreement with the prediction of 
gradient based models.~\cite{PBW-GGA}
This trend however does not carry over to low densities, particularly
in the atomic core [Fig.~\ref{nxcfig1}(c)].  The correlation hole here,
while clearly shorter-ranged than the LDA hole, is also significantly 
larger and makes a much larger relative contribution to $\exc$. 
The WDA and ADA both do moderately well in predicting the magnitude
and length scale of the correlation hole in this case.

In general the nonlocal density averaging methods fair less well at
predicting correlation or exchange alone than the combination of the two.
For the WDA in particular, the sum rules of the correlation hole 
and the exchange hole alone are not satisfied by satisfying the sum 
rule for exchange-correlation [Eq.~(\ref{eqwdasum})], with the correlation
hole influenced more by the short-range behavior of the angle-averaged
density and exchange the long-range behavior.  This may contribute
to the larger errors seen in $n_c$ and $n_x$
than for $n_{xc}$ observable in the cases with significant density variation.

%%%%%%%%%%%%%%%%%%%%% EPS INCLUDE FIG 2%%%%%%%%%%%%%%%%%%%%%%%%%%%%%
\begin{figure}[htb]
Fig. 2
\epsfbox{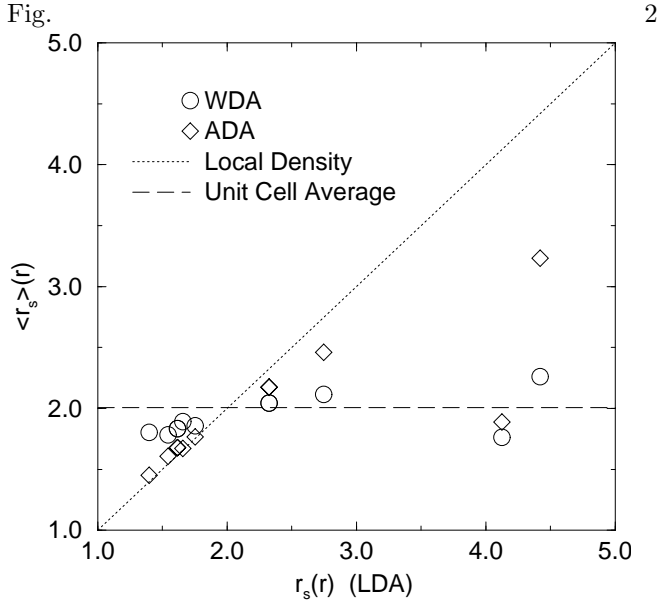}
\vskip 0.3cm
\caption
{Correlation lengths of WDA and ADA exchange-correlation holes.  The
correlation length for the equivalent homogeneous electron gas hole, 
$\bar{r}_s = ( 3/4\pi \bar{n}({\bf r}) )^{-1/3}$, obtained
from the nonlocal DFT models discussed in the paper is plotted versus the 
correlation length $r_s$ derived from the 
local density for holes at representative points throughout the unit cell. Dotted line gives the LDA approximation, dashed line the
$r_s$ factor obtained from using the unit cell average.
}
\label{nxcfig2}
\end{figure}
%%%%%%%%%%%%%%%%%%%%% EPS INCLUDE FIG 2%%%%%%%%%%%%%%%%%%%%%%%%%%%%%

Some sense of the general trends resulting from the different strategies
for choosing the pair correlation function
can be obtained from plotting the solution for the average interelectron
distance ${\bar r_s}$ used by the nonlocal methods versus the $r_s$ obtained
from the local density.  The result for sample points at various densities
throughout the unit cell is shown in Fig.~\ref{nxcfig2}.  
%The dotted line
%plotting ${\bar r_s}\! =\! r_s$ is the LDA approximation, diamonds give the 
%values for the ADA and circles those for the WDA.  In comparison, the 
%dashed line at $r_s=2.01$~a.u.\ is obtained from the unit-cell
%average density, the number of electrons per unit cell over the
%unit cell volume.  The high-density (minimum $r_s$) value of the ADA and WDA 
%data is that of the bond center and the two low-density outliers are 
%the atom center and interstitial point.
In the case of the ADA, for $r_s$ lower than the unit-cell average of
2.01~a.u., ${\bar r_s}$
hews closely to the local density value.  It deviates from it at lower
densities, either dramatically at the atom center (the one point
below the average $r_s$ curve), or more gradually as
one heads towards the interstitial point (the other low-density points 
plotted).  
%One may understand these results by recalling that the region over 
%which the ADA weighting function $w$ extends depends on $r_s$, while the 
%typical length scale required to probe significant differences in density 
%is on the order of the unit-cell average $r_s \sim 2.0$.  
The WDA values in contrast are all grouped about the unit-cell average 
value, regardless of location and density.  In other words, 
the variation in the WDA exchange-correlation hole 
is primarily derived from variation in the nonlocal angle-averaged
density, and very little from the variation of $\bar{g}$.  
%In addition, the 
%average ${\bar r_s}$ obtained in WDA apparently has little connection
%with the correlation length of the actual hole, unlike in the ADA.
These results reflect the character of the criteria used to determine the
average density in each case.  The weighting function of the ADA is
an oscillatory function in real space, with a peak contribution at
roughly 0.5~$r_s$ from the reference electron, so that a significant
effect is observed only where the density variation about the electron
is sufficiently rapid.
%and a node at 1~$r_s$.  The
%average density is determined mostly by the density variation within a 
%small volume about the reference electron, and is only significantly different
%from the local density when the density variation is very rapid.
In contrast, the weight implied by the WDA sum rule condition,
\mbox{$u^2 [1 - g^{heg} (u,r_s)]$}, is peaked near $r_s$ and has
significant contributions out to 2~$r_s$, a distribution broad enough to
to pick up the average density at almost every point in the crystal.
%and thus tends more towards 
%measuring the average density of the system.
%(NOTE that I can plot the VMC correlation length measured either as 
%the location of the minimum of the weighted, angle-averaged hole or
%as the cutoff radius inside which, say, 90$\%$ of the exchange-correlation
%energy is obtained.)  

%QUESTION:
%The data shown leaves unclear why ADA is a worse overall fit to our 
%best diffusion Monte Carlo estimate of $E_{xc}$ than
%WDA; both have basically the same $\exc$ in the points checked and the same 
%types of corrections over LDA.  (However, note how complicated the
%difference between each model and the VMC data actually is: one needs maybe
%10 points on the 111 axis to get a complete picture at high density.)
%The energy density differences plotted by 
%Randy at these points are on the order of 0.0004 to 0.0008 a.u.\  
%The resolution of the integrated
%holes that I plot is maybe .001 x n(r).  This is on average, .00003 a.u.\
%for the energy density.
%
\section{Exchange and correlation holes}
\label{results_bondcenter}
As discussed earlier, the trends in the comparison of the DFT and VMC models
for $\enxc$ are complex and difficult to characterize.
%Particularly, in the
%case of the WDA and ADA, these deviations form a
%complicated ``dapple" pattern with rapid variations within the Si unit cell.
Some insight into what is going on can be obtained by decomposing the
exchange-correlation hole into exchange and correlation 
components and particularly by considering the response not only as a function
of distance from the reference electron but including the complete information,
\mbox{$\nxc(\bfr,\bfr + \bfu)$}, of the second electron's 
position in the crystal. 

%(This may be an important conclusion . . . although only the spherically 
%averaged hole (in fact, only the first moment of this hole, as mentioned
%by Jones and Gunnarsson~\cite{JG}) is needed to obtain the energy density
%at a given point, it in itself offers only limited information about
%correlations.  In other words, the same reduction 
%of information by averaging
%that leads to the success of simple models like the LDA in modeling
%angle- and system-averaged holes also can hinder the understanding of what
%causes deviations from such models in realistic systems.)

The exchange hole is of use particularly in that it can be derived exactly 
and thus provide an unambiguous test of density functionals.  Further,
it is often the dominant part of the total exchange-correlation hole.
%REVISED BELOW:
%The exchange hole may be derived from
%Kohn-Sham orbitals provided that the Kohn-Sham equation
%reproduces the ground-state density of the fully interacting system,
%that is, by implementing the proper noninteracting limit specified in the
%adiabatic connection formula for $E_{xc}$.  
%In practice the LDA orbitals
%produce a density that is indistinguishable from the VMC density for
%bulk Si within the statistical limitations of the Monte Carlo sampling.
%By applying the Slater determinant ground-state wave function for $\lambda=0$
%into Eq.~\ref{eqnxc} one has
%\begin{equation}
%    n_x({\bf r},{\bf r}+{\bf u}) = 
%	- \frac{1}{n({\bf r})}\; \sum_{\sigma} \left| \sum_{\alpha}^{N_\sigma}
%	  \psi_{\alpha}({\bf r}) \psi_{\alpha}^{*}({\bf r}+{\bf u}) \right|^2.
%\end{equation}
%where $\sigma$ is a spin index, $N_{\sigma}$ the number of electrons with 
%a given spin, which is equal to the number of valence states periodic on the
%$3 \times 3 \times 3$ simulation cell.
It may be formally obtained as the exchange-correlation
hole associated with the Slater determinant ground-state wave function
of the noninteracting ($\lambda\!=\!0$) limit of Eq.~\ref{eqHam}.  
By applying this wave function 
in Eq.~\ref{eqnxc} one has
\begin{equation}
    n_x({\bf r},{\bf r}+{\bf u}) = 
	- \frac{1}{n({\bf r})}\; \sum_{\sigma} \left| \sum_{\alpha}^{N_\sigma}
	  \psi_{\alpha}({\bf r}) \psi_{\alpha}^{*}({\bf r}+{\bf u}) \right|^2,
\end{equation}
where $\sigma$ is a spin index and $N_{\sigma}$ the number of electrons with 
spin $\sigma$. 
In practice $n_x$ is obtained from LDA valence orbitals 
periodic on the $3 \times 3 \times 3$ simulation cell used in our VMC
calculation.
These produce a density that is indistinguishable from the VMC density 
within the statistical limitations of the Monte Carlo sampling.
At ${\bf u}=0$ the exchange hole reduces to $-n({\bf r})/2$, which reflects the
Pauli exclusion prohibiting the occupation of the same point in
space by two particles of the same spin. 
%or for a spin-unpolarized
%system half the electron density is removed by the exclusion principle.
In addition, the exchange hole is strictly negative, as may easily be seen
in the case of Si where the orbitals may be taken as real.

%%%%%%%%%%%%%%%%%%%%% EPS INCLUDE FIG 3, FIG 4 %%%%%%%%%%%%%%%%%%%%%%%%
\begin{figure*}[t]
%Fig. 3
\epsfbox{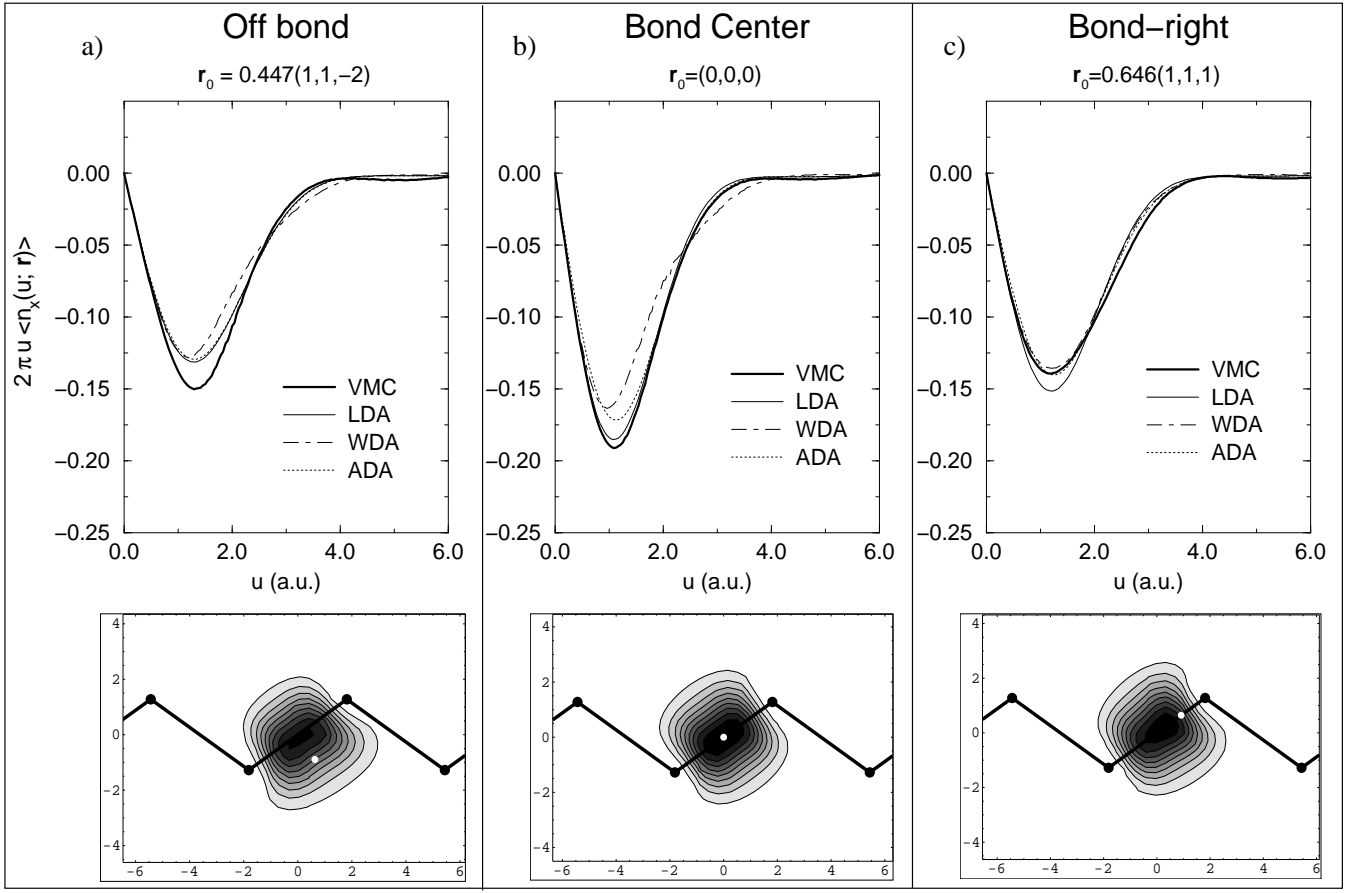}
\vskip 0.3cm
\widetext
\caption
{Exchange hole near bond center: (a) off bond, perpendicular to bond axis,
(b) bond center, (c) off bond center along [111] axis.  Top half shows
angle-averaged hole $\langle n_{x}(u,{\bf r}) \rangle$ weighted by $2\pi u$ 
versus distance from reference 
electron $u$ for VMC data and several models.  Bottom half shows
contour plot of $n_{x}( {\bf r}, {\bf r} + {\bf u} )$ in (110) plane, 
with location of reference electron as white dot.  Gray scale is in 
increments of 0.005~a.u.\ with white
region between -0.005 and 0.000~a.u.
}
\narrowtext
\label{nxcfig3}
\end{figure*}

\begin{figure*}[t]
%Fig. 4
\epsfbox{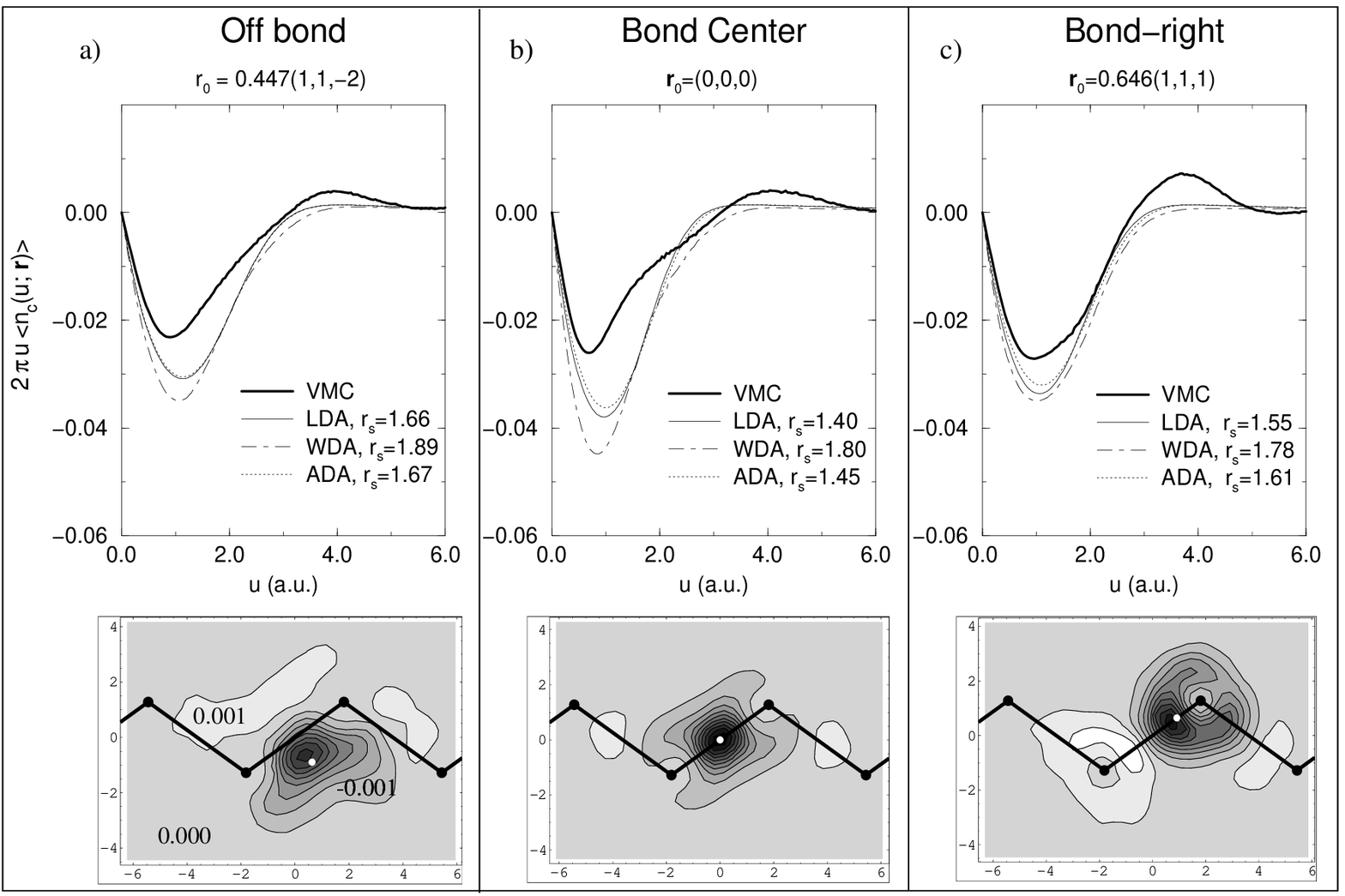}
\vskip 0.3cm
\widetext
\caption
{Correlation hole near bond center: (a) off bond, perpendicular to bond axis,
(b) bond center, (c) off bond center along [111] axis.  Top half shows
angle-averaged hole weighted by $2\pi u$ versus distance from reference 
electron $u$ for VMC data and several models .  Bottom half shows
contour plot of $n_{c}$ in (110) plane, with location of reference 
electron as white dot.  Gray scale is in increments of 0.001~a.u.\ with white
region between 0.001 and 0.002~a.u.
}
\narrowtext
\label{nxcfig4}
\end{figure*}
%%%%%%%%%%%%%%%%%%%%% EPS INCLUDE FIG 3, FIG 4 %%%%%%%%%%%%%%%%%%%%%%%%

In Fig.~\ref{nxcfig3} we show results for the exchange hole and angle-averaged
hole weighted by distance for three positions near the bond center of the
Si crystal.  
%The exchange hole is exactly determined from the LDA orbitals used to 
%create the Slater determinant component of our correlated trial wave function.
In the bottom half of the figure is a contour plot of
$n_x({\bf r}, {\bf r}+{\bf u})$: the exchange hole at ${\bf r} + {\bf u}$ 
given a reference electron at the position ${\bf r}$. 
The plots cut through 
a chain of Silicon atoms in the (110) plane of the crystal;
the reference electron's position is shown as a white dot.  The shading
represents a change in value ranging from just less than zero for the white
region to just above the absolute
minimum of minus one-half the peak density of 0.087 a.u.\ in 
the black regions.  
The thick solid curve in the upper plots show the result of spherically
averaging the exchange hole to obtain a function of distance from the 
reference point.
%averaging over the hole 
%at constant distance $u$ from the reference point ${{\bf r}}$.  
The same weighting factor of $2\pi u$ as that of Fig.~\ref{nxcfig1} is used. 
This result is compared to the various DFT models discussed previously.

In the case of the reference electron at bond center, the exchange hole
is focused in the bonding region immediately surrounding the reference
electron, with the bulk of the hole contained in the region between the
atoms on either side of the bond, and a negligible weight
on the nearest neighbor bonds.  
The electron density has been reduced by roughly 45$\%$ within the bond,
i.e., in a diamond-shaped region extending 1.7 a.u.\ along the bond axis
in either direction and 1.2 a.u.\ perpendicular to it.  This accounts for 
90$\%$ of the 
density of electrons with the same spin as the reference one.
%Greater 
%than 45$\%$ of the total electron density is removed within a diamond shaped 
%region extending 1.7~a.u.\ along the bond axis in either direction and 
%1.2~a.u.\ 
%perpendicular to the bond.  This is to be compared to a maximum of 50$\%$ of 
%the density that might be removed in principle by Pauli exclusion.  In other
%words, given a test electron at a bond center, 
%there is less than 10$\%$ probability of finding a second electron 
%of the same spin as the first within that bond.
%which induces spatial correlations only between particles with 
%the same spin.
As one moves the reference electron either along the bond axis (c) 
or perpendicular (a)
to it, the hole tends to remain focused on the nearby bond center, despite
a small shift in weight towards the reference point.
As before, more than 90$\%$ of the same-spin density is removed 
in the bond.  This stiffness or insensitivity to the position of the reference
electron is gradually lost as the reference electron is moved,
say, to the atom center or towards the antibond point equidistant from
three bonds.  In the latter case, the exchange hole is again centered
on the reference point and resembles a $sp^3$-hybrid atomic orbital.
%(cite Hybertsen and Louie possible).

The effect of the stiffness of the exchange hole on its angle average can
be seen in the difference between the LDA and exact cases (thin and
thick solid lines).  In the bond center, Fig.~\ref{nxcfig3}(b), the LDA result 
closely matches the exact value.  At
position (a) off the bond axis, the bond center is 
oriented tangentially with respect to an
angle average about the reference point and the exchange hole has become
deeper and narrower than that obtained by the LDA.  In contrast, in case (c), 
where the bond center is oriented radially out from the reference 
electron, and the electron is near to an atomic core, the hole has 
become shallower
and wider than in the LDA.  Although the two positions have similar
densities and density gradients, the resulting deviations from the LDA are
qualitatively different and point to a genuinely nonlocal behavior in 
the exchange hole.  Nonetheless,
neither the WDA nor the ADA are particularly
sensitive to the position of the bond center with respect to the
reference electron and provide no systematic
improvement (or disimprovement) over the LDA fit.  

The correlation holes corresponding to each case in Fig.~\ref{nxcfig3} are
shown in Fig.~\ref{nxcfig4}.  Note the order of magnitude
smaller range of density changes involved.  The correlation hole $n_c$ 
takes on both
negative values in regions where electrons are repulsed
by the Coulomb potential of the reference electron and positive values
representing regions where electron density has increased.  The net
electron density change is zero.  
%For the contour plots, the grey region in 
%the background represents zero change in density, lighter areas represent the
%region of density enhancement, and darker those of density reduction.

At the bond center, Fig.~\ref{nxcfig4}(b), the 
correlation hole contour plot can be roughly separated into two regimes: 
a deep and narrow
well in the vicinity of the reference electron and a shallow 
longer-ranged response that may be described as a polarization of the nearest 
neighbor bonds.  In the region of the pseudopotential core the correlation 
response
is suppressed.  The resulting angle-averaged hole has 
a complex and unusual structure as a function of distance. 
%which is in general narrower and weaker than predicted by the LDA.
%The minimum of the 
%distance-weighted hole is both weaker and shifted considerably towards the
%reference electron as compared to the LDA hole.  
%The narrow well about this minimum 
%broadens considerably at longer distances, extending out to 
%a width comparable to that of the exchange hole.
%The electron density removed from the vicinity of the reference electron
%is localized about a peak at 4.0~a.u.\ that is correlated with the peaks 
%in $n_c$
%on the six nearest-neighbor bonds, as shown in the contour plot.
In comparison, the LDA and other DFT holes show a relative lack of 
structure and greatly overestimate the magnitude of the correlation
energy per particle at bond center.
%One can compare this to the relative lack of structure in the LDA or other
%DFT holes, which also

As the position of the reference electron is moved
either off the bond axis (a) or along the axis (c), the correlation hole
demonstrates a marked sensitivity to its position relative to the bond center.
In either case the minimum is slightly offset in the direction of 
steepest increase in density.
This is consistent with $g_c$ being nearly isotropic near the electron 
as expected from the cusp condition.  Outside this short-range region
the hole undergoes strong distortions in shape from that 
of the hole about the bond center.  These
longer wavelength features, as in the bond-center case, are
well characterized in terms of the polarization of the nearby bonds.  
Bond polarization is particularly noticeable for the electron fixed on the
bond axis~(c) where there is a large shift of the electron density 
from the atom nearest the electron to the other side of the bond, with
a peak on the bond axis opposite the reference electron.
Nearest neighbor bonds are polarized in a similar fashion. 
In (a), the nearby bonds are polarized with density shifted
from the nearer side of the bond with respect to the electron position
to the farther, with the node of the polarization approximately along the 
bond axis.
%In (a), the electron density is reduced in a broad area
%behind the fixed electron with respect to the closest bond
%center, and enhanced in a similarly broad region on the side of the 
%bond opposite the reference electron. 
%%The electron density of the nearest-neighbor bond is similarly polarized.
%%again, with a reduction in density on the near side of the bond and 
%%an enhancement of the density on the far side.  
%The polarization of the bond
%is even more apparent for the case in which an electron is fixed
%on the bond axis to the right of bond center~(c).  There is a considerable
%reduction of electron density in a region extending through the nearest atom
%and a considerable enhancement of density on the atom on the other side
%of the bond, with a peak on the bond axis opposite the reference electron.
%In both cases, nearest neighbor bonds are also polarized, 
%roughly along 
%the line joining the reference electron and the bond center, 
%with a node along the bond axis in case (a) and perpendicular to it in (c).
Consideration of the corresponding pair correlation function shows that these 
features in the correlation hole constitute a significant
departure from an isotropic form of pair correlation function such 
as is assumed by the DFT models discussed in this paper.

% COMPLETE REDRAFT 01/01/01 !!!!!!!!!!!!!!!!!!!!!
Despite the reduction of information engendered by averaging over angle,
the dramatic changes in the shape of $n_c$ with position show up in the
angle-averaged hole, giving rise to nonlocal corrections to the LDA hole.
The LDA model depicts a trend in $n_c$ to a broader and 
shallower hole as the density decreases from case (b) to (c) to (a) in
the upper half of Fig.~\ref{nxcfig4}.  The VMC hole in case (a) retains
the character of the bond center hole, being narrower and weaker than that
predicted by the LDA.  The overall disagreement with LDA is 
less pronounced and some of the detailed structure is lost.  
As one moves along the bond axis to case (c), however, a significantly
different trend occurs: the hole in the region of density reduction
becomes deeper and broader relative to the LDA, and in particular now
matches the width of the LDA hole.  In addition to the stronger response
in this region, a large peak appears at the hole edge in keeping with the
zero sum rule of the correlation hole.  
This peak, at about 3.5~a.u., correlates with the 
position of electron density peaks on the nearest neighbor bonds in the 
contour plot.  
As with the case of exchange, these trends in $n_c$ poorly correlate
to the changes in density or density gradient, and show  
a sensitivity to the larger nonlocal crystal environment.
%The effects of bond polarization features in the correlation hole 
%plays a noticeable role in its spherical average, despite the 
%considerable reduction in information engendered by the average over angle.
%When the electron is moved off the bond axis, the weighted hole 
%retains the general character of the hole at the bond center, 
%although the disagreement with LDA is somewhat less pronounced and some 
%of the detailed structure is lost.  
%%weakness and the inward shift of its minimum relative to the LDA apparent in
%%the hole at bond center.  
%The correlation hole on the bond axis (c) is at short range much
%closer to the LDA hole in shape and magnitude than the other
%two cases and has an unusually large peak at the 
%edge of the hole.  
%This peak, at roughly 3.5~a.u., correlates with the 
%position of electron density peaks on the nearest neighbor bonds in the 
%contour plot.  
%%The strong peak on the bond axis is cancelled out by the 
%%strong reduction in density behind the nearest atom.  
%As with the case of exchange, then, correlation holes with similar local 
%densities and gradients show a sensitivity to the larger nonlocal crystal environment.

Though the exchange and correlation holes both show significant and 
nonlocal deviations
from the isotropic form typical of the homogeneous electron gas, these
deviations are highly correlated with each other and thus
tend to cancel in their sum.
%The insensitivity of the exchange hole near the bond center with respect to
%the reference electron position is countered by a marked sensitivity 
%to reference position in correlation.  
As the exchange hole tends to stay on the bond center, the
bond polarization in the correlation hole is oriented so as to shift the 
center of the exchange-correlation hole towards the reference
electron.  
A simple picture of this feature in the correlation hole
is that it represents a response to the anisotropic
distribution of the exchange hole 
by which some of the deviations of the exchange hole from an 
isotropic form are screened out in the correlation hole.
This behavior may help to explain the relative success of the LDA in
describing the exchange-correlation hole. 

%A response of $n_{xc}$ to the position of the reference
%electron, and insensitivity to other details of the environment of the
%hole, is typical of the ``dynamic" correlation associated with the
%homogeneous electron gas, in contrast to ``nondynamic"
%features that reflect the electronic structure and are associated
%with orbital effects.  The cancellation of such nondynamic effects
%in exchange by a corresponding nondynamic component in correlation
%provides a reasonable explanation for the cancellation of the LDA
%errors in exchange and correlation that occur in their sum.

%%%%%%%%%%%%%%%%%%%% EPS INCLUDE TABLE 2 %%%%%%%%%%%%%%%%%%%%%%%%%%%
\begin{table*}[htb]
\widetext
\caption{Exchange and correlation energies $E_x$ and $E_c$
in eV per atom for VMC and various density functional methods
described in the text.  The results of a diffusion Monte Carlo DMC calculation
to remove the variational bias due to the use of a trial wave function is
also shown.
}\label{table2}
\begin{tabular}{l|rrrrrr}
 & LDA & ADA & WDA & GGA & VMC & DMC \\
\tableline
\mbox{$E_x$} & -27.66 & -27.56 & -27.37 & -29.10 & -29.15 & -29.15 \\    
\mbox{$E_c$} & -5.09 & -5.11 & -5.63 & -3.93 & -3.58 & -4.08 \\ 
\mbox{$E_{xc}$} & -32.75 & -32.67 & -33.00 & -33.03 & -32.73\mbox{$\pm$}0.01 
   & -33.23\mbox{$\pm$}0.08     
\end{tabular}
\narrowtext
\end{table*}
%%%%%%%%%%%%%%%%%%%% EPS INCLUDE TABLE 2 %%%%%%%%%%%%%%%%%%%%%%%%%%%

Neither the LDA nor the nonlocal DFT's model
with much fidelity the variation in the VMC correlation hole,
aside from the general broadening of the hole with lower density.
In particular the 
corrections of the WDA lead to a worse fit of both the correlation hole 
and its energy.  The differences in exchange and correlation
between WDA and VMC do tend to balance each other but not consistently:
for example they cancel nicely in~(a) and~(b) but add in case~(c).  
%The inability to model the subtle structural effects in exchange and
%correlation that we observe near the bond center
%may explain then the limited success of the WDA in predicting the
%high density behavior of the energy density as observed 
%previously.~\cite{Hood2}
The net effect of these trends on the exchange-correlation energy
is shown in Table~\ref{table2}, where the total $E_{xc}$, $E_x$ and $E_c$
for several methods are compared.  In addition to the methods discussed
in detail in this paper, we present results from the PW91 version of the
GGA~\cite{PW91} and those of a diffusion Monte Carlo (DMC) 
calculation.~\cite{CepChK,DMC}  The DMC method removes nearly
all of the variational bias in the VMC correlation energy at a considerably 
larger computational cost.  With the exception of the GGA,
none of the density functional methods obtain a good estimate for $E_x$ or
$E_c$ as compared to DMC, with errors of roughly 1 to 2 eV 
per atom.  Estimates for $E_{xc}$ are much closer, especially for the WDA.  

%(It is interesting to note that the GGA 
%calculation, using the PW91 functional~\cite{PW91} does a much better 
%job at describing $E_x$ and $E_c$.  One factor that may account for this
%is that the functional form for exchange-only structure function $S(k)$ in the
%homogeneous electron gas varies as $k$ for small wavenumber, leading
%to a long range tail in the pair correlation function.  This term is caused
%by the sharp cutoff in momentum in the momentum density at the
%fermi momentum and is cancelled out exactly by a similar
%term in the correlation hole.  (Note that the structure function
%of the homogeneous electron gas, including both exchange and correlation has
%a $k^2$ behavior at small $k$ due to plasmons.)  The
%effect on the correlation hole is that the electron density removed from 
%the immediate vicinity of the reference electron is spread out over the
%entire crystal in an anomalously slowly decaying fashion, and
%thus has minimal effect on the correlation energy.  In contrast the
%VMC data show that most of the redistributed density lies near
%the edge of the exchange hole, forming the well defined peak seen in 
%Fig.~\ref{nxcfig4} which contributes to a significant reduction in the
%energy.  Any exchange or correlation-only model based on the homogeneous 
%electron gas should suffer from a similar treatment of the long-range hole.
%On the other hand, the GGA form uses explicit cutoffs in both the exchange
%and correlation hole, which avoids this problem altogether.)  

\section{Discussion}
\label{Discussion}

\subsection{Limits of density averaging}
The clear success of density averaging occurs at low densities 
where ADA and WDA obtain excellent exchange-correlation energy densities.
WDA in particular predicts the shape of the exchange-correlation hole
with exceptional fidelity, even in the extreme situation inside the atomic
core.  At high density, the small level of variation in the nonlocal 
angle-averaged density [Eq.~(\ref{eqnav})] limits the effects of 
density averaging to subtle
alterations of the hole which do not provide a systematic improvement with 
respect to the LDA.  More importantly, the
discrepancy between the density-averaged holes and the VMC hole 
(much less the exact hole) are difficult to correlate with any known quantity.

However,
density averaging must lead to a global and systematic improvement over the
unit cell or else the quality of the total result may be disappointing.
This is particularly the case of the ADA, which returns values very close
to the VMC values at low density and slightly underestimates the magnitude
of $E_{xc}$ at high density.  Unfortunately, once the variational bias
of the VMC energy is removed by a DMC calculation as shown in 
Table~\ref{table2} this result proves to be a significant disimprovement 
with respect to the LDA.  Moreover, although the different prescriptions for 
the WDA and LDA holes lead to extreme differences in the degree
of nonlocality they incorporate,
%or a minimal versus a large sensitivity to the local density,  
both end up with quite similar, reasonable predictions for the total 
$E_{xc}$. 
%in contrast
%to the ADA that has something of an intermediate form.  
%Thus
%the marked difference in the shape and length scale of $n_{xc}$ accounts
%for a relatively small change in energy.  
Apparently, the large error at low densities
of replacing $n({\bf r} + {\bf u} )$ with the 
local density $n({\bf r})$ in the definition of the LDA exchange-correlation 
hole is offset by correspondingly large changes in $g$, which are suppressed
by density averaging in the WDA.  
%The LDA hole is that of an actual physical 
%system, which guarantees a number of exact 
%as well as energetically reasonable approximate properties of $n_{xc}$, 
%and as such is far more robust than might otherwise be expected.  

It is interesting to compare the density averaging approach to the 
gradient expansion used as the basis for the GGA.  
A notable result of the PW91 version of the GGA is that it not 
only provides a significant
improvement of the LDA exchange-correlation energy, but of the exchange
and correlation energies as well (Table~\ref{table2}).  Thus, it alone 
among the functionals we have studied shows promise to be a reasonable
candidate as a correlation-only density functional for this material.  
In comparison, the WDA, although it returns a very good $E_{xc}$, has
the worst estimate for $E_c$.  
%However,
%the small-distance expansion of <n> in terms of the local density and its 
%Laplacian which is  the formal equivalent of the information about the 
%environment used in generating the GGA hole, 
%is only a small fraction of the information potentially usable in the WDA.
However, the local density and its 
Laplacian, the formal equivalent of the information about the surroundings
of an electron used in 
obtaining a GGA hole, constitutes but a small part of what is contained in
the angle-averaged density $\langle n({\bf r},u)\rangle$. 
The limitation of the WDA is the restriction to a 
pair correlation function form with only one variable parameter, one to 
which the hole in our case proves largely insensitive.  It would be 
interesting if more of the information contained in 
$\langle n({\bf r},u)\rangle$ could be used as input
into a more flexible model for $g$, particularly for correlation or
exchange separately.  Possibly, the rate at which the
density changes over the length scale of the hole, a factor which,
as illustrated in Fig.~\ref{nxcfig1}, is clearly
important in determining the overall shape of the WDA hole,
could be used to influence the form of $g$ in analogy to the GGA.

Furthermore, $\langle n({\bf r},u)\rangle$ has truly nonlocal
information not accessible to GGA's.  A salient feature of the system
in our study is 
the long-range limit of the angle-averaged density, which tends to a finite
constant for a crystal and to zero for an atom or molecule.  
It is thus possible that 
low density points in these two systems have similar local or semilocal 
environments (and thus the same GGA holes) but  
significantly different exchange-correlation holes because of 
different long-range boundary conditions.  
Because of the nonlocal character of the WDA, these boundary conditions 
have a strong influence on the WDA hole, causing the length scale $\bar{r}_s$ 
to be nearly constant for the crystal but for the atom, to range  
from a finite value near the valence density maximum to infinity as 
the reference electron is removed from the atom.
%condition leads to the invariance of the WDA hole with respect to 
%the scaling length $\bar{r}_s$ 
%for the Si crystal, in contrast to a significant variation 
%in $\bar{r}_s$ in the Si atom.
The WDA produces a reasonable fit to $n_{xc}$ at low density for
both the Si atom and Si crystal, despite the 
markedly different $\bar{r}_s$ factors necessary to satisfy the 
particle sum rule.~\cite{Puzder}  
It would seem, then, that the WDA should be capable
to match or surpass the quality of the GGA -- perhaps given 
a more accurate and flexible model pair correlation function form.

%The exchange-correlation
%hole developed as a generator for the PBE exchange-correlation energy 
%functional is not directly comparable to our data.
%The integration by parts used to eliminate the Laplacian of the density in 
%the gradient expansion leads to an altered $n_{xc}$, removing the significant 
%changes in $n_{xc}$ at points where the Laplacian is large, such as the atom 
%center, and transfering the effects onto the $n_{xc}$ at points with high 
%density gradients.
%One may attempt to compare qualitative features in the GGA hole 
%associated with rapid changes in density as measured by the gradient of the
%density $\nabla n({\bf r})/n({\bf r}) to the effects of density
%changes in our data where the 
%Laplacian or gradient is large.  In particular the GGA predicts the
%exchange hole to become deep and more short-ranged in an inhomogeneous
%system.  This is true of our data generally only at low density -- at
%high densities, although a long-range tail in the homogeneous electron
%gas hole (due to the sharp jump in the momentum density at the Fermi 
%surface) the remnant is typically shallower and more wide than the LDA.

\subsection{The exchange-correlation hole and energy density in the GGA}

As discussed in Sec.~\ref{theory_nxc}, the GGA 
models gradient expansion corrections to the LDA in terms of the gradient of
the density alone, mapping corrections proportional
to its Laplacian to a gradient squared term by an integration by parts
in Eq.~\ref{eqexc}.  
This transformation has significant implications for the GGA models
of $n_{xc}$.
For example,
each position in Fig.~\ref{nxcfig1} is a critical point in the density,
and as such characterizes the nature of $n_{xc}$ in
its vicinity.  At these points, the PBE model for $n_{xc}$ is 
indistinguishable from that of the LDA, a particularly serious error in
the atom core region~\ref{nxcfig1}(c), where effects of inhomogeneity
are most apparent.  At the same time, corrections to  $n_{xc}$ from 
the atom core and other regions where the Laplacian is large are 
mapped to the $n_{xc}$ of regions where the gradient is large, producing
misleading corrections in these regions as well.  
The GGA is rather designed to provide gradient corrections to the LDA on a 
system-averaged basis.
It does in fact capture general trends in exchange and correlation
as discussed briefly in the next section, but clearly not on
a local or point-by-point basis.
%The improvement
%of GGA over LDA is by construction restricted to the system-averaged
%hole or the integral of the $n_{xc}$ over reference position
%$\bfr$, and $E_{xc}$, and should not be 
%expected to be a guide for improving 
%local quantities such as $n_{xc}(u,\bfr)$ and $\epsilon_{xc}(\bfr)$.  

This approach is frequently justified by the observation that
the energy density is not in itself a necessary component of 
an accurate DFT, 
rather only the total $E_{xc}$ and its functional derivative with
respect to the density, the exchange-correlation potential.~\cite{JG}
Nonetheless, 
%the exchange-correlation
%potential has an important contribution from the exchange-correlation
%energy per particle,~\cite{footexc} and it is reasonable to assume tha 
%Moreover, in the Si crystal it proves 
it is worth mentioning that the deviation of the LDA energy 
density from the VMC energy density 
closely follows the 
Laplacian of the density.  For example, the minimum 
of the Laplacian, an hourglass-shaped region near the bond center,
closely correlates with the region where we have found the largest positive 
deviation of the LDA energy density from the VMC value.~\cite{Hood1}
Likewise a large negative deviation in the LDA energy
density occurs in the atom center where the Laplacian has a maximum,
and a smaller negative deviation in the interstitial region where the 
Laplacian has a weak local maximum.  This latter feature is clearly
visible in the \nxc and \exc data of Sec.~\ref{results_avnxc}.
We
expect that a GGA model that fits both gradient and Laplacian terms 
could prove very useful in improving the energy density and thus also 
the exchange-correlation potential.

\subsection{Orbital effects in exchange and correlation}
%Consistent with the particle sum rule that the exchange hole represents
%the removal of exactly one electron (the one fixed at the reference point)
%one may interpret the hole as the probability density of the orbital
%occupied by the fixed electron.  In the case of the exchange hole near the
%bond center, this orbital is basically a ``bonding" orbital, and in the
%antibond region, more of an ``atomic hybrid" orbital.

Insight into the comparison of our VMC results and those obtained from the
various models derived from the homogeneous electron gas may be obtained
from a consideration of the semiconductor environment.

%%%%%%%%%%%%%%%%% EPS INCLUDE FIG 5 %%%%%%%%%%%%%%%%%%%%%%%
\begin{figure*}[htb]
%Fig. 5
\epsfbox{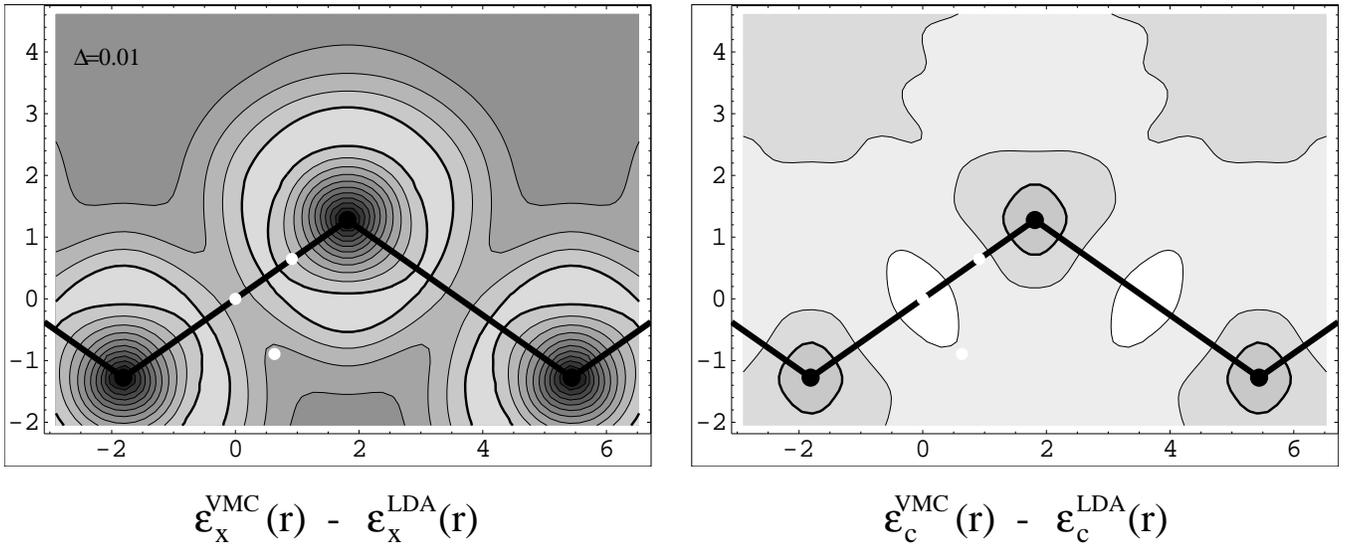}
\vskip 0.3cm
\widetext
\caption
{VMC exchange and correlation energy per particle in the (110) plane of the
Si crystal, relative to the LDA value.  The contours show increments
of 0.01~a.u., with the thicker contour showing where the VMC and LDA values
are equal.  White dots show the locations of the reference electrons about
which exchange and correlation holes are shown in 
Figs.~\protect\ref{nxcfig3} and \protect\ref{nxcfig4}.
}
\narrowtext

\label{nxcfig5}
\end{figure*}
%%%%%%%%%%%%%%%%% EPS INCLUDE FIG 5 %%%%%%%%%%%%%%%%%%%%%%%

First of all, a distinguishing feature of systems such as molecules or
semiconductors which have a finite energy gap is that the ground state
may be described in terms of an exponentially decaying localized basis.  
%In a periodic system these are the Wannier functions, 
%which are the maximally localized basis set derivable by a unitary 
%transformation from the valence Bloch orbitals.
%Wannier functions in Si can be defined by a unitary
In a periodic system these are the Wannier functions, 
which in Si can be defined by a unitary
transformation from the four valence bands to four orbitals 
per unit cell localized about
the bond centers ${\bfr_I}$:~\cite{DesC,Zak}  
\be
     W_I(\bfr - \bfr_I) = \sum_{nk} U_{nk,I} \psi_{nk} (\bfr).
\ee
The symmetry of the crystal requires that each orbital $W_I$ must be
related to the others by a space group operation.
Such orbitals are well localized on a given bond
and describe well the bonding character of the valence electrons.
The exchange
hole is most clearly represented in terms of these local functions and
deviates strongly from the homogeneous electron gas model when these 
functions have little overlap.
The exchange hole in terms of a Wannier basis is 
\bea
    \nonumber n_x & & ({\bf r},{\bf r}+{\bf u}) = \\
	& & - \frac{1}{n({\bf r})}\; \sum_{\sigma} \left| \sum_{I}^{N_\sigma}
	  W_I(\bfr - \bfr_I) W_I^{*}(\bfr+\bfu-\bfr_I) \right|^2.
\eea
Using Wannier functions obtained from our LDA orbitals and the method 
of Marzari and Vanderbilt,~\cite{MV} we find that 97$\%$ of the 
the on-top value $n_x(\bfr_I,\bfr_I)$ of the exchange hole at bond center
is determined solely from the 
Wannier function localized on that bond site. 
Thus, the lack of sensitivity of the exchange hole with respect to reference 
position seen in Fig.~\ref{nxcfig3} near the bond center is likely a 
reflection of the domination of the exchange hole in this region by a
single Wannier function.  This situation is 
reminiscent of that of the ${\rm H_2}$ molecule in which the exchange 
hole is 
constructed from a single orbital and is totally insensitive to electron
position.  

An exponentially decaying exchange hole typical
of an insulator or finite system has the effect of lowering the
exchange energy with respect to that of the LDA, owing to the
more localized form of the hole as compared to that of the homogeneous
electron gas; imposing a finite-ranged hole has been an ingredient
in constructing successful GGA methods.~\cite{PBW-GGA}  
In our case the actual $E_x$ is 1.5 eV lower than the LDA result, 
in good agreement with the PW91 GGA result; likewise, the
the exchange energy per particle $\ex$ is lower than the LDA prediction 
through much, but not all, of the unit 
cell as shown in Fig.~\ref{nxcfig5}.
%In comparison one may consider the ${\rm H_2}$ molecule in which
%the exchange hole is composed of essentially one orbital: the exchange hole
%is then totally insensitive to the position of the reference electron.

The semiconductor environment also affects the correlation hole
in several ways.  First, 
there is a finite energy cost to correlate electrons as compared to the 
homogeneous electron gas and therefore there should be a smaller correlation
response and correlation energy than predicted in the LDA. 
This is seen in the VMC correlation energy density
and energy per particle at almost all 
points in the unit cell, as shown in Fig.~\ref{nxcfig5}.
Again this trend is consistent with the assumptions made in the PW91 version
of the GGA but does not correlate with those made in the density 
averaging methods.
%(It is also a typical feature of covalently bonded molecules? cite )

Secondly, as the electronic structure of the covalent
bond plays an important role in exchange (in the form of a bond-orbital-like
exchange hole near the bond center), it should do so as well in
correlation.  The signature of the existence of the covalent bond 
in the correlation hole is bond polarization, a clear feature of our
VMC data.  In the simplest bond-orbital picture,~\cite{Harrison} 
this arises from the introduction into 
the noninteracting ground state 
%(characterized near bond center by the bond orbital-like exchange hole) 
of excited states in which a 
pair of electrons are excited from bonding to antibonding orbitals on 
the same bond site.~\cite{Fulde}
These states contribute a ``left-right" correlation similar to that
of ${\rm H_2}$ and other diatomic molecules in which the probability 
density of the second electron is shifted along the bond axis
to the opposite side of the bond as that occupied by the first electron.
A similar excitation of nearest neighbor electrons causes van der Waals
like correlation between bonds.~\cite{Fulde}  

These effects are evident in the
``bond-right" case (c) of Fig.~\ref{nxcfig4} where
the antibond orbital has a peak.  At the same time, the correlation hole
associated with this left-right polarization is negligible when the 
reference electron is placed on the nodal plane of the antibond which
passes through the bond center normal to the bond axis: that is, 
the situation of cases (a) and (b) of Fig.~\ref{nxcfig4}.  
Consequently, the weakness of $n_c$
relative to the homogenous electron gas based DFT models in these two 
cases and the enhancement of the correlation hole on the bond axis have
a plausible explanation in the differing contribution of the 
antibond correlation to the holes in each case.~\cite{foot1}  
%and apparently account for the enhancement of the correlation
%hole at this position as compared to the other two positions considered.
%The correlation energy density at a given point in the crystal should 
%depend on how much the antibond excitation contributes to 
%the correlation hole about that point.  
%In particular,
%the contribution to the correlation hole due to the 
%excitation of an electron pair from bonding to antibonding 
%orbitals within a single bond is negligible for a 
%reference point on the antibond node passing through the bond center 
%and has a maximum off the bond center along the bond axis, 
%where the antibond has a peak.
Moreover, bond polarization 
contributes noticeably to the functional variation of $\ec$ in this
bonding region.
We find that the $\ec$ obtained from
VMC data has a maximum (that is, a minimum in the energy reduction obtained
by electron correlations) on a narrow ridge centered on
the antibond node passing through the center of the bond 
with normal parallel to the bond axis.
Although some of the variation in $\ec$ in the bonding region can be
accounted for within the LDA, the behavior is sufficiently dissimilar to 
the local density variation as to lead to the largest
discrepancy in the unit cell between the VMC and LDA results, 
as shown in Fig.~\ref{nxcfig5}.  

The significance of the antibond correlation and bond polarization
in general is that it provides a partial
explanation of the cancellation of errors in the LDA exchange and 
correlation hole seen in their sum.
The exchange-correlation hole of the homogeneous electron gas is
essentially ``dynamic" in nature, that is, responding 
to the position of the reference electron and largely insensitive 
to the details of electronic structure.  The ``nondynamic" or orbital-dependent
features in the exchange hole, such as its insensitivity to  
particle position near bond center are a large potential source of error
for the LDA and other models based on the homogeneous electron gas.  The
partial cancellation of these effects by a corresponding nondynamic feature
of the correlation hole should then provide a combined hole much more
amenable to the LDA and related density gradient corrections.

We have made preliminary estimates
of the antibond contribution to the
correlation hole $n_c$ and the correlation energy per 
particle $\ec$ in the bonding region of Si 
using perturbation theory and perturbative configuration interaction 
methods~\cite{Fulde} and localized orbitals derived from pseudopotential
plane-wave DFT orbitals.  We reproduce the qualitative
differences in $\ec$ between the points studied in this paper,
as well as a reasonable reproduction of the bond polarization features of
Fig.~\ref{nxcfig4}.  Details of this calculation are to be reported in a 
later paper.

\subsection{Error analysis}
%The effects of plane-wave cutoff in $n_{xc}$ can be estimated by comparing the
%exact exchange hole and that obtained
%with a plane-wave expansion limited
%to the basis set used for obtaining the pair correlation function.
As discussed in Section~\ref{method}, the exchange-correlation hole suffers
from fairly large plane-wave cutoff errors in the atomic core where
the pseudopotential orbitals vary rapidly. 
%with neglible changes elsewhere.
In order to produce a reasonable correlation hole, particularly in the
core, it was necessary to correct for this error by taking the
$n_x$ and $n_{xc}$ expanded to the same cutoff.  Then a best estimate
for $n_{xc}$ was taken by adding the finite cutoff estimate of $n_c$ to 
the exact $n_x$, that is, an estimate of $n_{xc}$ obtained from correcting
the exchange component of the Monte Carlo estimate:
\begin{equation}
  n_{xc} = n_{xc}^{approx} + n_{x} - n_{x}^{approx}.
\end{equation}
The change in $n_{xc}$ is noticeable particularly for the atom
center case Fig.~\ref{nxcfig1}(c), where the plane-wave correction
results in a deeper and more localized $n_{xc}$.  
The resulting $\exc$, shown in Table~\ref{table1}, varies from the raw
VMC data by roughly $8\%$, increasing the 
agreement with WDA and ADA at the expense of the LDA value.  However,
given the poor agreement of the LDA and VMC values, the uncertainty
in the VMC value does not significantly alter the assessment of a dramatic
improvement by ADA and WDA in this region.
A rather smaller change was noticeable for the angle-averaged hole
for the ``bond-right" position discussed in Figs~\ref{nxcfig3} 
and~\ref{nxcfig4}.  In this case, the difference is noticeable on the scale
of the energy differences between the various theories, with the 
plane-wave correction favoring the ADA case.
The correction was essentially negligible for the other points studied.

A second source of error in the VMC data is due to the statistical nature
of the Monte Carlo estimates of expectations.  Statistical sampling
leads to roughly equal errors in the expectations of each measured plane-wave
component of the pair correlation function, and a homogenous
background noise (up to the resolution of the plane-wave expansion)
in the real space behavior of the function.  With the assumption of 
homogeneous and uncorrelated background noise, the weighted angle-averaged 
holes shown should have an error that is roughly constant at large 
interparticle distances.  We find this to be a reasonable description
of the long-range fluctuations in this quantity, resulting in 
a very small error for all the cases we have studied.  
On the other hand, the energy and particle
sum rule of the exchange-correlation hole, both integrals over a large volume,
suffer more serious cumulative effects from background noise,
particularly at large interparticle distances where $n_{xc}$ is 
essentially zero.
As a result, the VMC particle sum rule as estimated from the plots in 
Fig.~\ref{nxcfig1} typically varies by 1 to 3$\%$ from the correct value.
%Most of this error comes from background fluctuations at large
%interparticle distances where $n_{xc}$ is essentially zero.
The values for the exchange-correlation energy per particle and energy density 
reported in our previous studies~\cite{Hood1,Hood2} have
been made by adjusting the exchange-correlation hole at each $\lambda$
to obtain the correct particle sum rule, and numerical values at the
various points considered in detail here are shown in Table~\ref{table1}.
This correction turns out to be noticeable at high density relative to the 
small differences between VMC and the various DFT models.

A final source of error is variational bias in the trial
ground-state wavefunction resulting from 
its limited variational flexibility.  The resulting discrepancy
from the true ground state can be significant for correlation: the
optimized VMC correlation energy is 14$\%$ higher than that estimated
by the nearly exact DMC calculation.  Exchange, on the other hand, is
relatively unaffected by variational bias, as the VMC density and 
the associated single-particle orbitals are expected to be very accurate.

Convergence studies~\cite{Cancio,Cancio2} of a Slater-Jastrow trial 
wavefunction for the Si atom
reveal trends that we also expect to be observed in the present study.
The correlation hole averaged over the position of the reference electron
should be quite close to that of the true ground state, with deeper
minima and sharper maximum in proportion to the change in $E_c$.  The
hole as a function of position in the system will in addition show
subtle alterations in shape, while preserving basic qualitative features,
for example the bond polarization in Fig.~\ref{nxcfig4}.  

%The choice of variational parameters by energy or energy variance 
%minimization favors to some degree the optimization of the wavefunction 
%for configurations with large values of the many-body wavefunction
%$\left| \psi_{\lambda}\right|^2$ at the expence of those with smaller
%values.  Hence we expect holes at high density to be better specified
%than those at low density, and that at the bond center to be specified 
%perhaps even to accuracy greater than the average correlation energy.
%Nevertheless, in the Si atom the shape and magnitude of the correlation
%hole at low density was generally at least crudely obtainable, perhaps
%because of the satisfaction of the cusp condition in the wavefunction.
The correlation hole can be expected to be more accurate at higher densities
which carry greater weight in a variational optimization of the wavefunction;
however the Slater-Jastrow wavefunction guarantees important conditions
on the hole at any density, such as the cusp condition and sum rules, 
so a reasonable
estimate of the hole should be obtained even at quite low densities.
Assuming that the generic trend at high density is to deepen the 
correlation energy of the hole by roughly $14\%$, 
we expect the overestimate of \ec by the DFT models, as shown 
for example in Fig.~\ref{nxcfig5} to be reduced by between 10 and 50$\%$
in the Si bond.  Qualitative trends in this region will not 
be changed if the error in \ec does not vary dramatically
with position.

In general, the various errors in the 
VMC data (plane wave cutoffs, sum rule enforcement, and 
variational bias) are most significant for the angle-averaged $n_{xc}$ and
exchange-correlaction energy per particle 
at high density.  Even though estimated corrections are rather small, 
the agreement with the various DFT's studied is quite close and even
small changes in the VMC data can be significant.  
Consequently, the errors in our data contribute to the general 
difficulty we find in assessing the high density trends of the WDA and ADA.
On the other hand, the difference 
between the DFT theories and the VMC are larger for the angle-averaged
correlation hole.
%that leads to the closer agreement in the exchange-correlation hole.  
The qualitative trends in $n_c$ near bond center discussed in this paper are
accordingly unaffected by error corrections though there are 
small differences in the quantitative value of $\ec$ at the
``bond-right" position.  Finally, the improvement of 
ADA and WDA over LDA at low density is so dramatic that the observed
error corrections to the VMC are small in comparison.  

\section{Conclusion}
\label{Conclusion}

We have carried out a detailed analysis of the exchange-correlation hole
and energy per particle in the Si crystal, comparing various DFT 
models to accurate numerical data calculated with the VMC method.

We find that 
the WDA and ADA help overcome the major defects of the local density
approximation at low densities and especially in the pseudopotential atom core,
where the rapid change in density relative to the length scale of the
LDA hole dramatically affects the shape and range of the hole,
and thus improve the fit with the VMC energy density.  
The remaining discrepancies are due to the failure of density averaging to 
provide significant information at intermediate and high densities,
where the inhomogeneity in the density is less severe, but the contribution
to the energy density of subtle effects of the inhomogeneity is noticeable
because of the higher density.
These discrepancies are, at least in principle, the result of the 
inflexibility of the scaling form of the pair correlation function 
used to fit the hole, which is insensitive to subtle variations in density.  

The detailed investigation of exchange and correlation at high density
reveal the importance of orbital correlations in both cases.
We find that the exchange hole is well described 
by a Wannier bonding state near 
the bond center, and that a ``left-right" correlation or bond polarization 
plays an important role in the 
spherically-averaged correlation hole and has a noticeable effect on the
correlation energy density.  This bond polarization 
has the effect of countering the coarse-scaled deviation of the exchange
hole from an isotropic form centered on the electron.  
As a result,
the exchange-correlation hole and energy density is much more reliably fit by
the DFT models studied than either exchange or correlation alone.

Viewed in another sense, these orbital effects lead
to serious defects in the correlation hole and energy of models derived 
from the homogeneous electron gas.
The WDA and ADA conspicuously fail to improve
on the LDA both in the system averaged energies $E_x$ and $E_c$ and in
detail, particularly in the bonding region.  Here, the nonlocal 
corrections to the LDA employed by these models
fail to account for the physical features responsible for the
most significant errors in the LDA model.

Of all the methods considered
in this paper, the GGA alone accurately predicts $E_x$ and $E_c$ for Si. 
Despite the lack of a direct comparison between our
data and current GGA models for $n_{xc}$,~\cite{PBW-GGA} the 
assumptions made in constructing the GGA exchange and correlation hole 
appear to be generally born out by our results, with a notable exception in the 
low density, large inhomogeneity limit for the correlation hole.
Thus we expect that the accurate GGA results for $E_x$ and $E_c$ are due
to an improved description of $n_x$ and $n_c$ on average, if not locally.
%We expect that the improved description of the averaged
%$n_x$ and $n_c$ to be responsible for the success of the GGA
%in obtaining $E_x$ and $E_c$.  
Interestingly,
the point by point trends of our data seem to be best 
characterized by the Laplacian of the density, in contrast to the approach 
taken by current
gradient-only GGA models.  Moreover, exchange and correlation taken alone
show nonlocal qualitative features at high density that may prove
difficult to describe in terms of any local density expansion.

It would be of great value with
respect to augmenting exact-exchange calculations to have a compact 
expression of the bond polarization features of the correlation hole in
terms of localized atomic or bond-centered orbitals, possibly with a 
short-range correction in LDA or GGA.
We have done preliminary
calculations on modeling this bond polarization in terms of excitations
of electrons into antibonding states.  However a conveniently usable and
compact form for the correlation hole at long range remains to be found.  

This work was supported in part by the Department of Energy
(Grant No. DE-FG02-97ER45632)
and the National Science Foundation (Grant No. DMR-9724303).  It was 
performed under the auspices of the U.S. Department of Energy by the
University of California, Lawrence Livermore National Laboratory under
contract No. W-7405-Eng-48.
We wish
to thank Nicola Marzari for his kind help in providing us Si Wannier
orbitals.


\begin{thebibliography}{99}
\bibitem{foot1} It is interesting to note that
even in the antibond-unfavorable cases there exist smaller polarization
features, either on nearest-neighbor bonds (b) or along a different
nodal plane (a).  These are not so easily explained in a simple bond-orbital
model but perhaps represent higher order orbital effects.
\end{thebibliography}
\end{document}